# Pioneering Advancements of 2D Graphene: Energy and Electronics applications.

Abdulrhman M. Alaraj [a,b*], Aya A. Esmaeil [a,b], Mohamed A. Khamis [a,b], Naglaa M. Zian [a,b], Fatma Sameh [a,b], Abdallah M. Abdeldaiem [a,b], Ebrahem H. Abdelaal [a,b], Walid J. Hamouda [a,b], Sara R. Ghazal [a,b], Habiba E. El-Sayegh [a,b], Aml F. Dawood [a,b], Fayza R. Ramadan [a,b], Haneen A. Saad [a,b], Mena. K. Selema [a,b], Nora. H. El-Mowafy [a,b], Mohamed M Kedra [a,b], Ahmed k. Abozaid [a,b], Ahmed M. Ragab [a,b], Ahmed A. El-Naggar [a,b], Walid Ismail [a,b], Swellam W.Sharshir [c], Abdelhamid El-Shaer [a,b], Mahmoud Abdelfatah [a,b*]

a Physics Department, Faculty of Science, Kafrelsheikh University, Kafrelsheikh,33516, Egypt
b Nano Science and Technology Program, Faculty of Science, Kafrelsheikh University, Kafrelsheikh, 33516, Egypt.
c Mechanical Engineering Department, Faculty of Engineering, Kafrelsheikh University, Kafrelsheikh 33516, Egypt.
\* Corresponding authors E-mail address: abdulrahman.mahmoud_a693@sci.kfs.edu.eg; mahmoud.abdelfatah@sci.kfs.edu.eg

**Abstract**

 This review explores the synthesis, characterization, and potential applications of graphene, a two-dimensional material with exceptional properties. Graphene's versatility in energy and electronics applications is highlighted, with its high conductivity and huge surface area facilitating improved energy storage capabilities in supercapacitors and batteries. In electronics, graphene is revolutionizing the industry by enabling the development of flexible displays, high-speed transistors, and enhanced thermal management systems. The integration of graphene into composite materials presents opportunities for stronger, lighter, and more conductive materials. The study provides a comprehensive overview of graphene's current and future impact on technology, emphasizing its transformative potential in energy solutions and electronic advancements. In the energy sector, graphene's integration into batteries, energy storage systems, capacitors, fuel cells, and renewable energy technologies signifies a leap forward in efficiency, capacity, and sustainability. In the electronics sector, graphene's unique characteristics are utilized in RFID, sensors, and EMI shielding, leading to communication, security, and device miniaturization advancements. The study underscores graphene's potential to spearhead future innovations, reinforcing its status as a pivotal material in the ongoing technological evolution.

| ABBREVIATION | WORD OR PHRASE |
|---|---|
| 0D | Zero-dimensional materials |
| 1D | One-dimensional materials |
| 2D | Two-dimensional materials |
| 3D | Three-dimensional materials |
| CVD | Chemical vapor deposition |
| QDS | Quantum dots |
| RFID | Radio-frequency identification |
| H-BN | Hexagonal boron nitride |
| TMDCS | Transition metal dichalcogenides |
| G | Graphene |
| RGO | Reduced graphene oxide |
| GO | Graphene oxide |
| GQDS | Graphene quantum dots |
| GNP | Graphene nanoplatelets |
| ROS | Reactive oxygen species |
| GICS | Graphite-intercalation-compounds |
| GNSS | Graphene nanosheets |
| GNWS | Graphene nano walls |
| GNFS | Graphene nanofibers |
| LIB | Li-ion batteries |
| CNT | Carbon nanotube |
| PEMFC | Polymer electrolytes for the membranes of fuel cells |
| ITO | Indium tin oxide |
| DSSC | Dye-sensitized solar cells |
| EMI-SE | Electromagnetic interference shielding |
| GAS | Graphene aerogel |
| HPF | Hydro plastic foaming |
| EP | Epoxy resin |
| PGFS | Porous graphene films |
| MGFS | Multilayer graphene films |
| RFID | Radio-frequency identification |

## 1. Introduction

In the pantheon of materials, few have sparked as much intrigue and promise as graphene. This two-dimensional wonder, a monolayer of carbon atoms arranged in a hexagonal lattice, has journeyed from theoretical curiosity to a cornerstone of modern material science. The genesis of graphene can be traced back to the early 20th century, but it wasn't until 2004 that researchers were able to isolate and explore its remarkable properties, a feat that earned them the Nobel Prize in Physics.

This two-dimensional layer, one carbon atom thick and arranged in a honeycomb-shaped lattice, has revolutionized our understanding of living organisms and fundamental physics [1-3]. Graphene's tapestry is woven with superlatives: it is the thinnest material known, yet incredibly strong, surpassing the might of steel. Its electrons dance at breakneck speeds, bestowing it with exceptional electrical conductivity, while its thermal conductivity is unparalleled. These attributes heralded a new era of potential applications, igniting research fervor worldwide [4, 5]. It is also impermeable to gases, highly flexible, and has a large surface area, making it an ideal material for various applications [6, 7].

Graphene fabrication techniques have developed rapidly, with methods for instance mechanical exfoliation [8, 9], epitaxial , and chemical vapor deposition (CVD)growth yielding high-quality samples [1]. These processes have been optimized to control the number of layers, crystal structure, and purity of the graphene produced, which are critical factors that determine its suitability for different applications [10, 11].

The importance of graphene extends beyond its impressive physical properties; it is a beacon of sustainability. As the world grapples with the energy crisis, graphene emerges as a key player in energy applications [12, 13]. Its high surface area and conductive prowess make it an ideal candidate for batteries, enhancing charge rates and energy capacity. In energy storage, graphene supercapacitors offer rapid charging and discharging capabilities, while in fuel cells, it improves efficiency and durability. The renewable energy sector benefits from graphene's lightweight and flexible nature, finding use in solar cells and wind turbines [2, 14, 15].

In the realm of electronics, Graphene's high electronic mobility makes it a potential replacement for silicon in transistors, leading to faster and more energy-efficient devices [2, 16]. Its transparency and conductivity [17] are used in the development of flexible touch screens, lighting panels, and solar cells [12, 13]. Its mechanical strength and flexibility are used to create new composites for stronger and lighter materials [18], particularly in aerospace and automotive industries [19]. Graphene's versatility in electronics includes making RFID tags more reliable and sensitive, enhancing the responsiveness of sensors, and shielding against electromagnetic interference.

However, the path to commercialization of graphene-based technologies is fraught with challenges, such as production scalability, cost effectiveness, and integration with existing materials and systems [20]. Despite these challenges, ongoing research and development into graphene synthesis and its applications continue to break new ground, indicating that the full latent of graphene has yet to be realized.

## 2. Bibliographic Evaluation

Bibliometric analysis is a method used to detect global research trends in specific fields by analyzing academic databases. It helps identify evolutionary characteristics and develop sections within a field, the number of published research from 2010 to 2025. As shown in **Fig. 1,** the pie chart, we detected that the most important part of graphene is used in energy storage areas and electronics, so graphene is a driver of modern technology applications [20, 21]

The rise in graphene publications in journals, as depicted in **Fig. 2**, underscores the significance of graphene beyond mere innovation. This article provides a detailed analysis of graphene's crystal structure, properties, fabrication techniques, and its applications in electrical and energy fields. It explores the history of graphene, its significance, and its numerous applications, highlighting its potential as a symbol of human ingenuity and innovation in the upcoming new age.

As shown in **Fig. 3**, the titles of the practical domains and the extent to which the articles published in them are related to other articles in the substance of the topic covered below. It is also shown in **Fig. 4**, which depicts a network representation of research connected to the research's keywords.

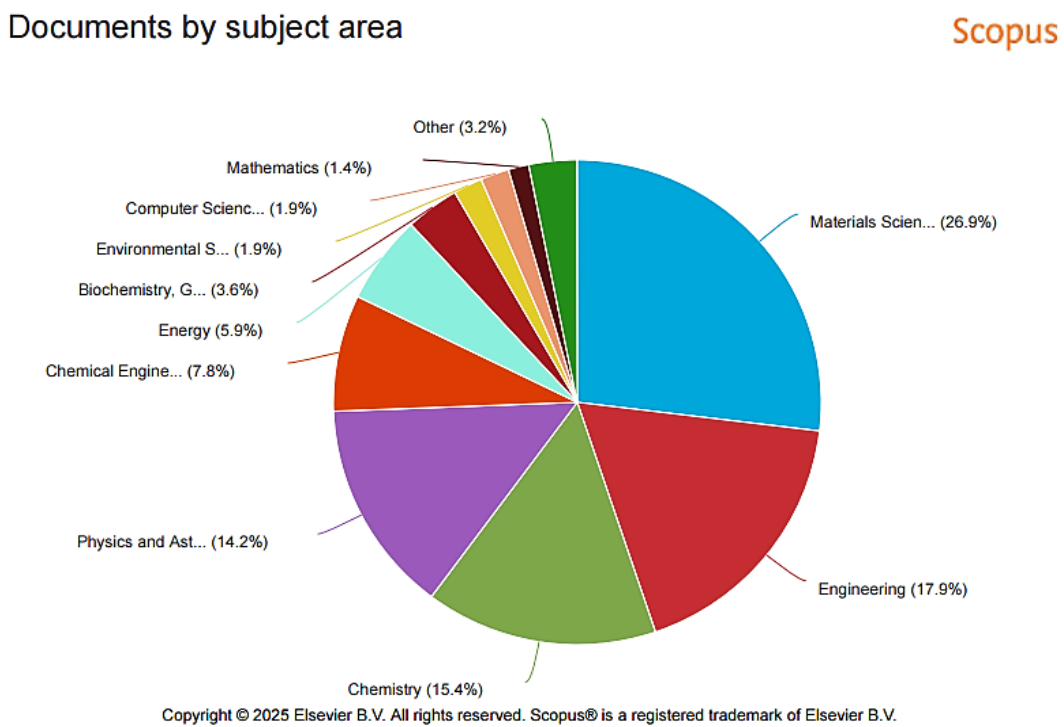

**Fig. 1**. Graphene-based industrial applications vary across various manufacturing regions.

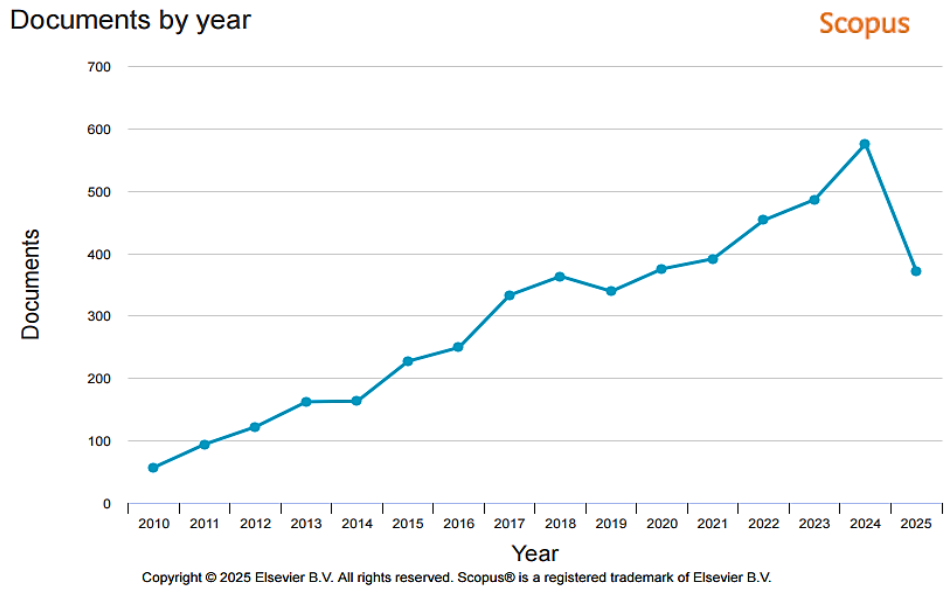

**Fig. 2**. The annual number of graphene research publications by years, Elsevier

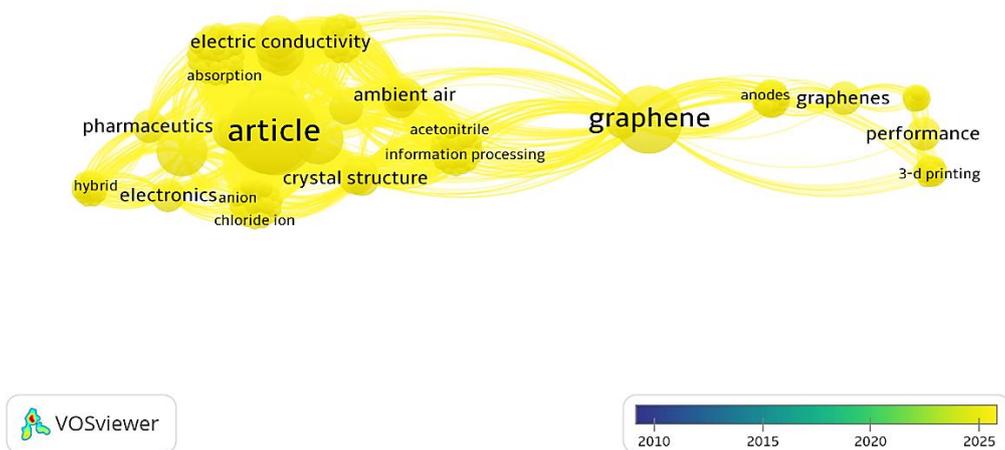

**Fig. 3**. Example of co-word (co-occurrence) overlay visualization using VOS viewer.

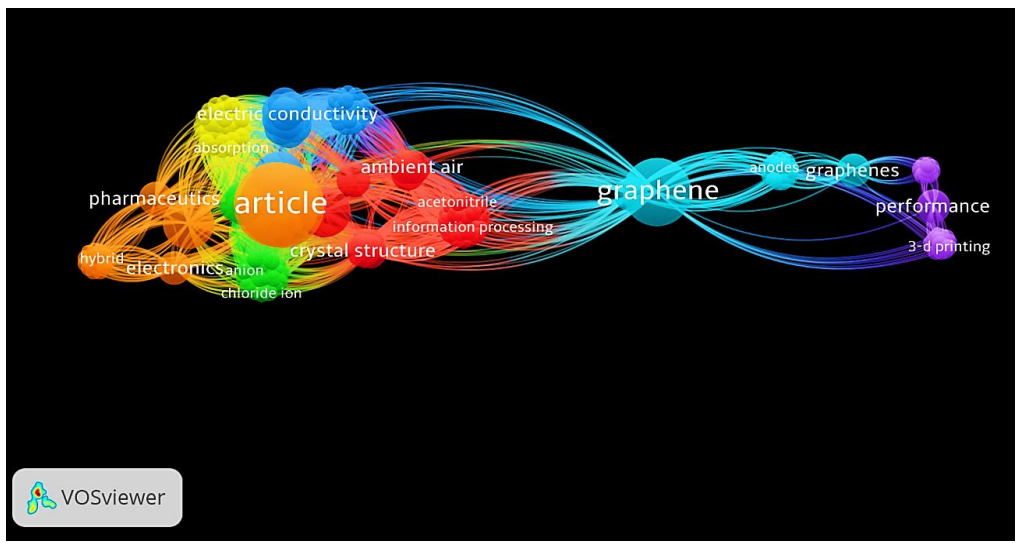

**Fig. 4**. Example of co-word (co-occurrence) network visualization using VOS viewer.

## 3. Classification of nanomaterials
### 3.1 Nanomaterials Based on Dimension

Nanomaterials are extremely small, having at least one dimension of 100 nm or less. 0-D, 1-D, 2-D, and 3-D are the four size-based categories into which nanomaterials are divided. They may be round, tubular, irregularly shaped, or solitary, fused, aggregated, or agglomerated, as shown in **Fig. 5**.

*Zero-dimensional nanomaterials* are substances that either have all their dimensions inside the nanometric range (>10 nm) or are completely confined inside the nanoscale range (x, y, and z). Among them are QDs, fullerenes, and nanoparticles.

*One-dimensional nanomaterials* have 2-D that fall inside the nanoscale range, but the other one is outside the range (>10 nm). One-dimensional nanomaterials encompass several forms as nanowires, thin films, nanofibers, nanotubes, and nano horns.

*Two-dimensional nanomaterials* have a 1-D (x) ranging from 1 to 100 nm, whereas two dimensions feature plate-like structures outside of the nanoscale. Nanosheets or walls, coatings, ultrafine grained over layers, tubes, fibers, free particles, and platelets are a few examples of two-dimensional nanomaterials. They are deposited on a substrate or integrated into the surrounding matrix material. They can be metallic, polymeric, amorphous, or crystalline.

*Three-dimensional nanomaterials* are substances that do not have a limit in any size or range of dimensions to the nanoscale. Dispersed nanoparticles, bundles of nanowires and nanotubes, and multi-nano layers with interfaces created by close contact between 0-D, 1-D, and 2-D structural elements make up their composition. Three-dimensional nanomaterials include colloids, thin films with atomic-scale porosity, and free nanoparticles with various morphologies. [22-26].

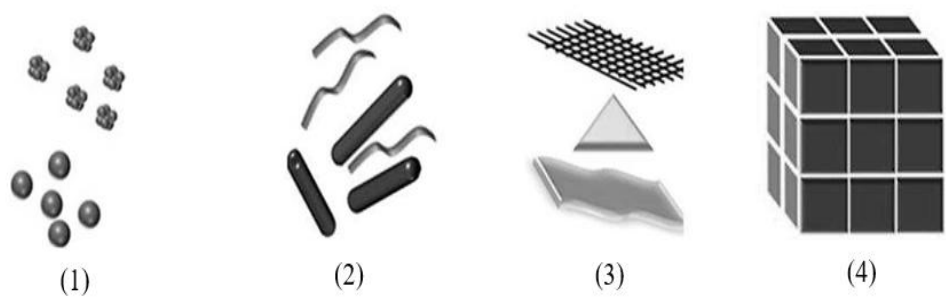

**Fig. 5**. (1) 0-D Spheres and Clusters, (2) 1-D Nanofibers, Wires, and Rods, (3) 2-D Films, Plates, and Networks, (4) 3-D Nanomaterials [24].

## 3.2 Quantum confinement effect

Quantum confinement refers to the spatially restricted energy states of electrons in nanocrystals, which become quantized when the material's spatial dimensions approach the de Broglie wavelength of its electrons. This results in distinct energy states, such as quantum dots in nanocrystals, which significantly alter the material's mechanical, optical, and electrical properties. Nanomaterials exhibit higher electron energies compared to bulk materials due to this unique quantum behavior. The energy gap within quantum dots increases with decreasing size, which is observable in semiconductor materials as their dimensionality reduces from two-dimensional to one-dimensional or zero-dimensional forms shown in **Fig. 6**, [27-30].

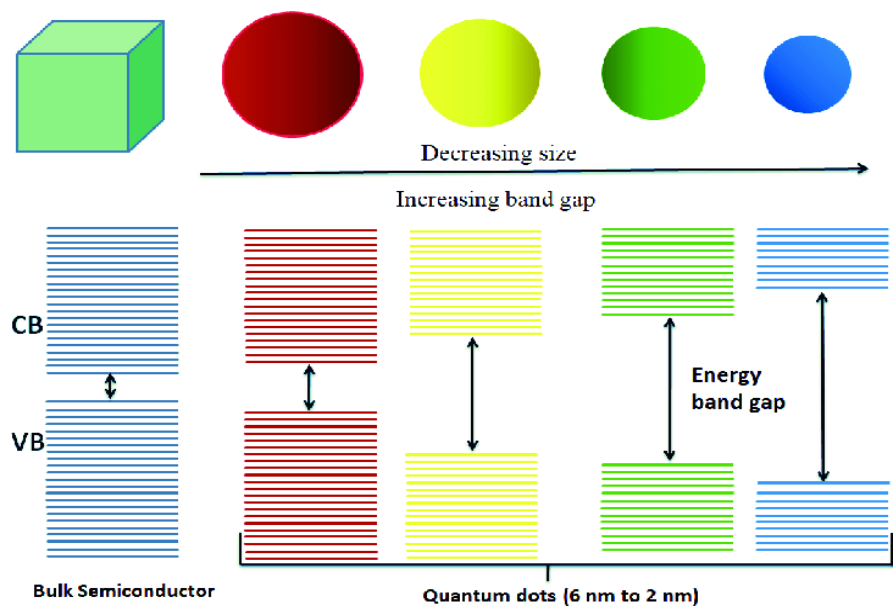

**Fig. 6**. The diagram illustrates the energy band structures in macroscopic bulk semiconductors and quantum-scale nanostructures, highlighting the distinct energy levels characteristic of quantum dots [30].

Quantum-confined structures are categorized based on their degree of confinement shown in **Fig. 7**. 2-D materials, like thin films and quantum wells, are confined along one dimension, while one-dimensional materials, like quantum wires, are confined across two dimensions. Zero-dimensional materials, like quantum dots, experience confinement in all three dimensions. Quantum dots display discrete energy levels and a distinct density of states, differentiating them from bulk materials and those with less dimensional confinement. The specific size of a material's quantum confinement varies and is influenced by its electron and hole mechanical properties [31, 32]. Recent decades have seen a surge in the development of nanostructures exhibiting quantum confinement, which have begun to revolutionize various applications. Among these, Quantum wires, along with QDs, stand out as promising contenders for advancements in lighting technologies and photonics applications [33, 34]. Also, QDs originate from the III–V or II–VI group elements of the periodic table and they are particles with dimensions smaller than the Bohr exciton radius, which is the threshold found in their bulk counterparts. As a result, pronounced effects of quantum confinement become evident. [35-37].

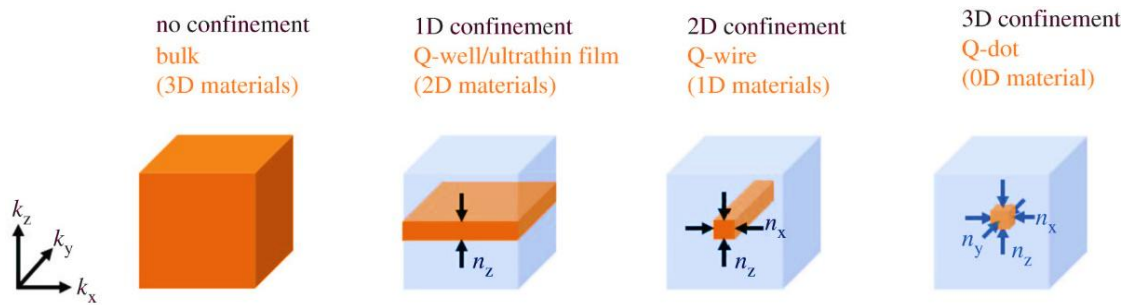

**Fig. 7**. The diagram illustrates the impact of spatial restriction in 1D, 0D, and 2D NMs, and the absence of such confinement in 3D bulk materials [31].

### 3.3 2-D Materials

Two-dimensional nanomaterials (2-D) have become a bright research point for scientists in the development of scientific research due to their distinctive chemical and electronic properties due to the confinement of the electron in two dimensions. Graphene is the most important two-dimensional nanomaterials that has been used for years due to its many advantages such as high specific surface area, quantum hall effect, high carrier mobility and great optical, electronic, and thermal properties [38]. Due to these advantages, graphene is used in many applications, including (energy storage and generation [39], hybrid materials[40],optical devices and high-speed electronics [41], chemical sensors [42], and DNA sequencing [43].

There are many nanomaterials that have this feature, such as (Transition Metal Dichalcogenides, Hexagonal Boron Nitride, Black Phosphorus, and Graphitic Carbon Nitride) [38]. These materials show diversity properties due to the advantages of their structure close to each other, but their structure is different from graphene. Two-dimensional materials with three-dimensional crystals stacked on each other, following van der Waals law, are crucial in applications like spintronics, catalysts, chemical and biological sensors, solar cells, supercapacitors, optoelectronics, and lithium-ion batteries due to their high surface area and importance in various fields.[38]

### *3.3.1 Crystal Structures Of 2-D Materials*

Currently, 2D materials, prepared through various industrial methods, can be classified into layered and non-layered types. Layered materials have strong chemical bonds between atoms, while non-layered materials are formed by van der Waals forces. There are numerous layered materials to mention.[38]

### *3.3.2 Graphene*

Graphene is a two-dimensional material made of one atom of graphite extracted from carbon and forming a closed hexagonal network of carbon atoms. The carbon atoms have 1.42 angstroms, and the single layers are stacked together by van der Waals forces, with each layer having 3.35 angstroms. Its unique properties include magnetic, electronic, optical, thermal, and mechanical properties, and it has a large surface area due to van der Waals forces stacking the single layers together, as shown in **Fig. 8** [44, 45].

### *3.3.3 Hexagonal Boron Nitride*

It is a layered graphite-type structure material and consists of planar meshes of BN-stacked hexagonal planes with a wide band gap (~5.9 eV). Contain (h-BN) With characteristic properties such as high temperature stability, thermal conductivity, mechanical strength, large and low dielectric constant. Due to the quality of its electrons, it has received wide attention in many applications [46].

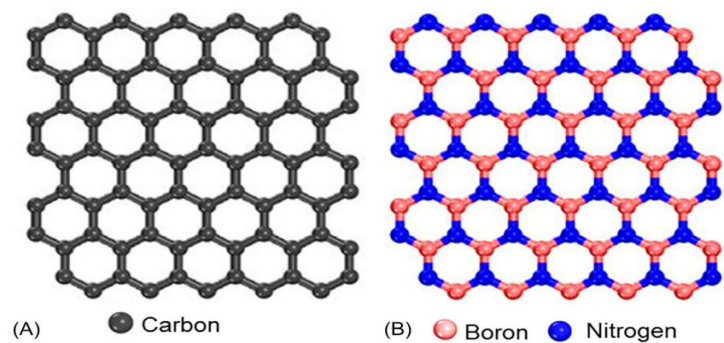

**Fig. 8**. (A) graphene and (B) hexagonal boron nitride (h-BN) [44, 45].

### **3.3.4 Transition Metal Dichalcogenides**

It's a new type of material and has properties that make it important for basic physical studies and for new applications such as nanoelectronics, nanophotonic, sensing and operation at the nanoscale. Interest in these materials has returned significantly since the presentation of the first transistor and also the discovery of strong photoluminescence in the monolayer.[47]

The materials consist of a hexagonal layer of metal atoms (M) sandwiched between two layers of chalcogen atoms (X). Each layer is typically 6-7 A thick, and it is two layers of chalcogen and consists of hexagonal metal atoms sandwiched between those two layers, as shown in **Fig. 9 (A)**. The MDX bonds within the layer are mostly covalent in nature, besides the van der Waals weak strength as well. This also makes it easier for the crystal to adhere to the surface of the layer. Metal atoms in TMDCs provide four electrons for bonding states, with chalcogen and metal atoms having oxidation states of 2 and +4, with bond lengths varying between 3.15 and 4.03 A, and layers can be prismatic, triangular, or octahedral, as shown in **Fig. 9 (B and C)** [48].

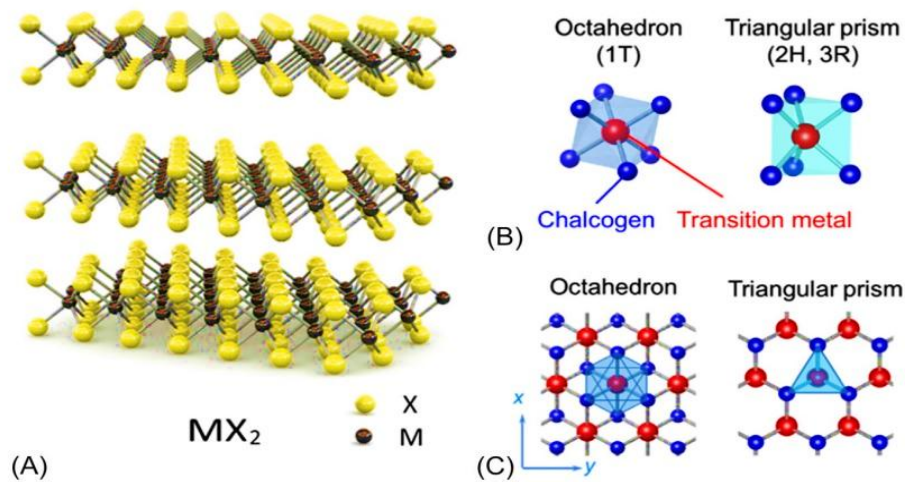

**Fig. 9**. (a) The 3D representation of transition metal dichalcogenides (TMDCs), (b) reveals octahedral coordination forming 1T structures and ternary prismatic coordination forming 2H or 3R structures, (c) with monolayers generated from this coordination [48].

**3.3.5 Black Phosphorus**

Black Phosphorus is another two-dimensional material that has become an important material. Black phosphorus crystallizes into a clear structure with perpendicular layers, held together by the van der Waals force, as shown in **Fig. 10 (a).** The monolayer, also known as phosphorene, consists of a vertically staggered hexagonal "armchair" structure with one P piece linked to the other three. The top view shows a hexagonal structure with conduction angles of 96.3° and 102.1°, as shown in **Fig. 10 (b).** Black Phosphorus exhibits a tunable bandgap ranging from 0.33 eV for the bulk to 1.5 eV for the monolayer, with a carrier capacity of up to 1000 cm2 V 1 s1. Due to its natural band gap and carrier mobility, black phosphorus could potentially bridge the gap between graphene and TMDCs [49].

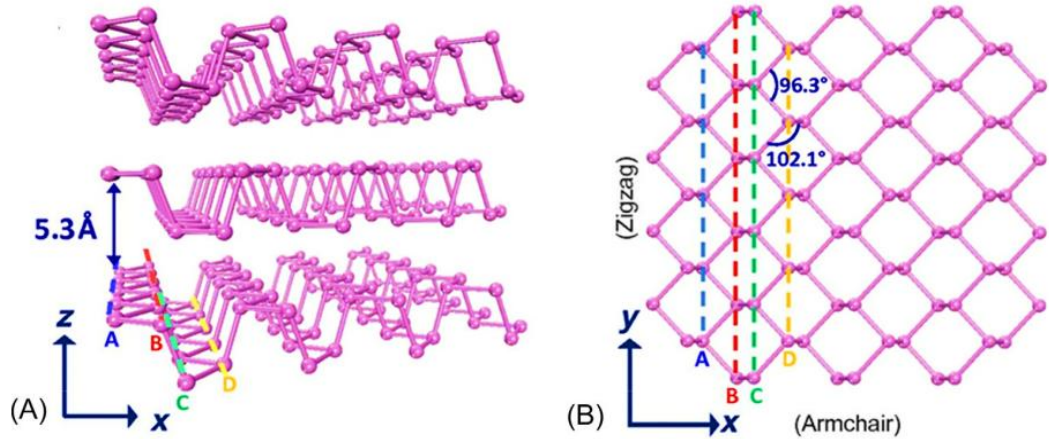

**Fig. 10**. (A) Side view of the Black Phosphorus crystal lattice. There is a divergence of the interlayer by 5.3A˚. (B) Top view of the lattice of single-layer Black Phosphorus. The bond angles are shown. The corresponding x, y, and z directions are indicated in both (A) and (B). x and y correspond to the armchair and zigzag directions of Black Phosphorus, respectively [49].

## 4. 2-D graphene and its types based on shape
### 4.1 Insight into graphene.

Due to its fascinating features, graphene has been the focus of extensive study over the last two to three decennary, among other 2D materials. The exploration of graphene significantly improved materials in science and nanotechnology, opening new routes. Applications of graphene's multidisciplinary characteristics extended widely, from aerospace [50] to health [51]. The focus of current graphene research has been on discovering novel graphene derivatives and using them in the creation of goods and machinery. Another novel strategy is to modify the surface or functionalize graphene to improve its characteristics [52]. Graphene is a newly discovered nanomaterial consisting of a 2-D $sp^2$ hybridized carbon atom honeycomb lattice. The important qualities of this material, including its large electrical conductivity, high thermal conductivity, strong mechanical properties, great chemical stability, and large surface area, have gained a huge importance in the fields of basic science and applied research. The most well-known graphene nanomaterials are mono and multilayer graphene, reduced graphene oxide (rGO), and graphene oxide (GO). These members of the graphene family have been used to research in various fields, including storage of energy [53], biosensing [54], nanoelectronics [55], catalysis [56], nanocomposites, and medicines [57], because of their exceptional qualities [58].

### 4.2 Graphene Types

While we've talked about graphene as a material it's worth noting that there exist types of graphene each dependent on the process of production. Graphene can be broadly categorized into groups; graphene, SiC graphene, exfoliated graphene, CVD graphene, Monolayer graphene, Graphene oxide, Graphene Nanoplatelets, few layers graphene or Multi-layer Graphene and reduced graphene oxide , as shown in **Fig. 11**. Unlike other types that come in sheet forms, reduced graphene oxide is commonly found in powder form. Can easily be  dissipated in water or other polar detergent  to create a paste.

#### a) *Exfoliated graphene*

 is produced by separating layers from a block of graphite. This was the first experimental method that succeeded in forming graphene in 2005. Although graphene may be found anywhere on a substrate and is just a few hundred μm in size, it is possible to generate almost perfect graphene with high crystallinity. Furthermore, exfoliated graphene remains limited  in terms of yield, size, and thickness control [59, 60].

#### b) *SiC and CVD graphene*

However, heating anything that contains carbon, such an organic molecule, will yield graphene, a carbide. Techniques utilizing this technique include SiC and CVD graphene. In other words, by evaporating silicon (Si) of SiC, a layered crystal of Si and C, at high temperatures, SiC graphene is created to precipitate carbon (C) atoms[61]. On the other hand, CVD graphene is produced by evaporating methane ($CH_4$) gas to deposit a carbon film on a metal substrate [62]. Between graphene and the Si substrate lies a C-rich layer, and graphene shows significant substrate interaction, despite the fact that the single crystalline size of SiC graphene is big and rather homogeneous, dependent on the initial SiC. The poor affinity of CVD graphene with the metal substrate allows it to be readily pulled off and separated from it, allowing it to be manufactured in bigger sheets than SiC graphene. The fact that it is created by depositing C atoms on a substrate means that the homogeneity isn't always high [63].

#### c) *Monolayer graphene*

its a mono atomic layer of graphite with a vast surface area characterized by a network of π electrons. Its exceptional intrinsic mobility [64, 65]. and high mechanical properties are attributed to its impressive Young's modulus. Single layer of graphene has favorable optical transmittance [65, 66] and thermal conductivity, making it useful in fields like field-effect transistors, transparent conductive films, sensors, nanocomposites, and clean energy devices [65, 67, 68].

### d) A few layers graphene (FLG) or Multi-layer Graphene (MLG)

consist of a small number (between two and ten) of precisely defined, quantifiable, stacked graphene layers with long lateral dimensions. These structures can take the form of sheet-like, flakes , free-standing films, or coatings bound to a substrate. They are valuable additions to composite materials, serving as mechanical reinforcements [1, 69, 70].

### e) Graphene oxide (GO)

an oxidized form of graphene with hydrophilic functional groups, offers easy solution processability in large quantities at low costs. Its primary advantage is its use as an additive in nanocomposite material fabrication, where monolayer graphene cannot be used directly [71]. Graphene oxide is a cost-effective and rapid material with a 2 dimention structure like graphene. It has a mono layer of carbon atoms covalently functionalized with oxygen-containing groups, making it easily processable in dispersions. GO's properties can be further customized through chemical engineering [72], making it suitable for various applications such as solar cells [73], sensors [74], supercapacitors, mcellular migration, neuronal generation, drug delivery [75], membranes, water purification, and multifunctional gels [76] [77].

### f) Reduced Graphene Oxide (rGO)

It's a form of graphene that has its oxygen concentration reduced through various methods, including chemicals and thermal treatment. It has structural flaws and is created from GO through a reduction process [78]. The process involves eliminating oxygen-containing functional groups [79] using environmentally friendly methods like hydrothermal [80], solvothermal [81], flash reduction [82], photocatalytic [83], and chemical approaches [80-84]. Hydrothermal rGO has favorable properties like porosity and structural stability, making it suitable for energy storage [85]. The rapid heating process decomposes oxygen-containing groups, making it an effective strategy for mass production of graphene in bulk quantities [86-88].

### g) Graphene Nanoplatelets (GNP)

It is a blend of monolayer, a few layers, and nanostructured graphite, with thicknesses ranging from 0.34 to 100 nm. They are lightweight, have, thermal conductivity and a high aspect ratio and offer excellent electrical, mechanical resilience, affordability, and a flat structure, making them an attractive alternative to other nanostructured fillers in materials science [89-91]. Graphene nanoplatelets are gaining popularity in various technological fields due to their enhanced tribological performance [92-94], mechanical strength [95-100], biomedical applications [101-104], gas barrier properties [105, 106], flame retardancy [107, 108], and heat conduction capabilities [109-112]. They can also enhance electrical conductivity on plastics [113-115], making them suitable for electronics and improving thermal conductivity of polymer matrices [116], making them effective thermal interface materials [117-120].

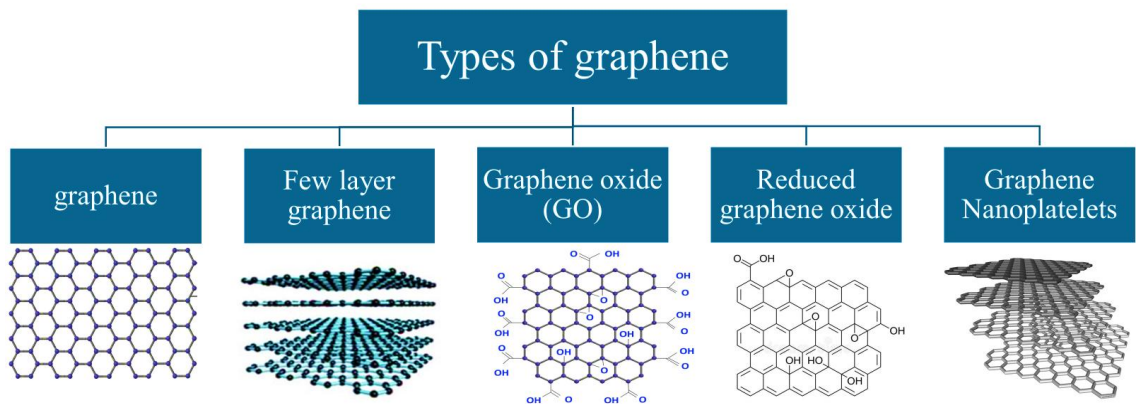

Fig. 11. Types of graphene based on shape.

## 5. Crystal Structure of graphene (G)

Graphene is a 2-D carbon allotrope [121, 122]. It consists of carbon atoms organized in a hexagonal design [123, 124]. A monolayer of carbon atoms arranged in such a honeycomb structure forms a mono graphene layer [125]. In graphene, all carbon atoms are covalently bonded to three other carbon atoms. In materials science, there are several allotropes of carbon identified counting diamond, fullerite, graphite, etc. As shown in **Fig. 12** [126]. Graphene, the novice It can be referred to as the strongest carbon allege as well as the best electricity conductor ever known [127, 128].

The bonds between carbon atoms in graphene, nanotubes, and buckyballs are crucial at the atomic level [129]. Also, Graphene is an allotrope of carbon with a 0.142 nm molecule length and graphite [130], formed by stacking graphene sheets, with an in-plane spacing of 0.335 nm [131].

Graphene has a homogeneous sp2 bond [132], and a rigid hexagonal backbone structure with strong in-plane bonds and out-of-plane bonds controlling interaction between different sheets. [133-135]. Anyhow, Changes in structure are mainly due to the absence of sp2 carbon atoms or the presence of sp3 hybridization atoms [136, 137].

Graphene layers stack to form graphite with an interplanetary distance of 0.335 nm (3.35 Å), and solid-state graphene sheets typically show diffraction evidence of (002) graphite layers. As shown in **Table. 1** [138] [139].

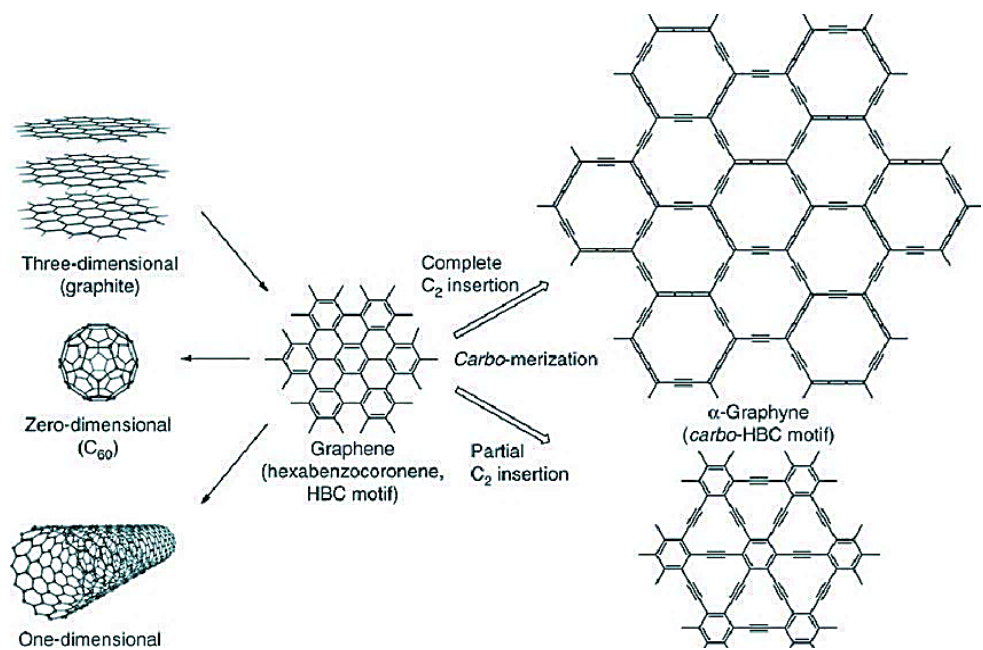

**Fig. 12**. structure of graphene(G) [140].

**Table. 1** Some basic physics constants of graphene were also summarized [141-143].

| Properties | Parameters |
|---|---|
| *Basic chemistry* | C :$(1s)^2(2s)^2(2p)^2$ |
| *Lattice structure* | Hexagonal lattice (Honeycomb) |
| *Lattice constant* | a=b=2.46 Å |
| *Interplanar spacing* | 3.37 Å |
| *planar spacing* | 0.335 nm. |

## 6. Properties graphene & Toxicity of GO

Nano materials based on graphene can be discussed depending on different aspects like mechanical, electrical, and impermeability. They also have thermal and optical properties. As shown in **Table. 2**.

### 6.1 Mechanical Properties

Graphene based nano materials are composed of planar sheets having dimensions in 2 planes. Those sheets consist of $sp^2$ carbon atoms connected with covalent bonds which are arranged in hexagonal patterns. Graphene exhibits a tendency towards brittleness, in addition to displaying non-linear elastic characteristics [144]. The non-linear elasticity observed when graphene oxide is subjected to a tensile load can be elucidated from the following equation: $\sigma=E\varepsilon+D\varepsilon^2$, wherein σ represents the stress applied to the surface (second Piola-Kirchhoff tensor) and E denotes Young's modulus. D stands for the third-order elastic constant which is empirically determined and ε signifies the elastic stress-strain. Specifically, E almost equals 1.0 TPa and D almost equals 2.0 TPa [66]. Brittle fracture phenomena manifest when a critical stress of a value 130 GPa is exerted along with a Young's Modulus of a value 1 TPa as per its intrinsic strength [66]. The substantial values of E and σint contribute to graphene being acknowledged as a material possessing remarkable strength suitable for structural purposes. Furthermore, the material's capacity to undergo bending represents an additional potentially advantageous attribute of graphene [145]. The computed average break strength and average elastic modulus for graphene oxide sheet reach values of 80 MPa and 32 GPa, respectively [146]. Additionally, graphene is distinguished by its remarkably low weight, measuring merely 0.77 mg/m$^2$. As per the Nobel announcement, a Graphene hammock scenario was depicted where area of 1 m$^2$ of graphene could uphold a cat weighing 4 kg, while weighing equivalent to just one of the feline's whiskers [69].

### 6.2 Electrical Properties

Graphene, an intrinsic semiconductor, exhibits a unique Fermi energy level configuration with a linear Dirac-like spectrum. Its electrical properties are highly regarded, with its overlapping bilayer band showing a negligible equivalent spectrum of 1.6 MeV [147]. The material is a metal-like substance with a unique exciton type, with complex band structures characterized by multiple charge carriers and overlap between valence and conduction bands, similar to graphene structures with three or more layers [148, 149]. The computed charge mobility of graphene on a non-conductive substrate is approximately, and $1.5 \times 10^4$ cm$^2$/Vs around $6 \times 10^4$ cm$^2$/Vs at temperatures of 27 C and -269 C respectively, with holes and electrons reaching a maximum concentration of $10^{13}$ cm$^2$ [148]. Therefore, owing to the exceptional conductivity of graphene specifically 30 Ω/sq cm$^2$, its sheet resistance is remarkably low [149, 150].

### 6.3 Optical Properties

Optical properties of graphene shows a high opaque atomic mono layer through vacuum [151]. It has a high luminescent property in addition to ability to absorb ( 2.3 %) from white light [151]. Additional layer of graphene can increase the quantity of absorbed white light [151]. Optical intensity reaches a certain limit and saturable absorption occurs, Due to graphene's property of ultrafast wavelength-insensitive absorption, Full band mode locking is achieved [151]. At the same value Graphene photograph (97.7%) of the visible light. The thickness can be increased by the linear reduction of graphene transmittance ,Go quantum dots photoluminescence and band gap able to be determined via the range of G-QDs which can emit light from blue color to red color ( 2.05 -2.9 ) eV [152, 153].

### 6.4 Thermal Properties

In graphene electronic thermal conduction is very small as the density of carriers graphene which is non-doped is low. So, at high temperatures thermal conductivity is considered as a phonon transport is comprising diffusive and in case of the minimal temperatures is ballistic conduction [152]. Long-range π-conjugation of graphene produces a distinct thermal behavior. It is determined to be high thermal conductivity for GO (up to $5\times10^3$ W/mK). The mentioned value is larger than the values of diamond and graphite. It can be contrasted with the conductivity of CNTs. It show that the strength of $sp^2$-hybridization C=C bonds realize efficiency of heat transfer through phonon transport models and lattice vibrations [154].

### 6.5 Impermeability

Graphene is a highly permeable material with high strength, densely packed carbon atoms, and large electron density. Its permeability has been used in applications like selective gas permeation, liquid containment for wet electron microscopy, ultrathin composite coatings for rust prevention, nanopore bio-diffusion devices, and molecular permeability preservation. GO also has impermeable laminations, high strength, and high electron density [155, 156].

**Table. 2** Summary of several main properties of graphene(G).

| | | |
|---|---|---|
| Mechanical | • Graphene is rather stiff.<br>• non-linear elastic behavior<br>• considered as a very strong material.<br>• light | [66, 144] |
| Electrical | • intrinsic semiconductor<br>• distinctive arrangement of the Fermi energy level<br>• linear Dirac-similar spectrum<br>• low sheet resistance i. e. 30 Ω/sq $cm^2$ | [147-150] |
| Impermeability | • high impermeability<br>• passing electrons through very greatly | [155, 156] |
| Optical | • high opaque atomic mono layer<br>• Ability to absorb (2.3%) of the white light.<br>• Full Band Mode Locking<br>• transparency (97.7%) for visible light | [151-153] |
| Thermal | • Insignificant thermal conductivity ($5\times10^3$ W/mK)<br>• distinct thermal behavior | [154] |

### 6.6 Toxicity of graphene:

Graphene nanomaterial is extensively utilized across various disciplines, particularly in the realm of biomedical applications owing to its distinct physicochemical characteristics. Numerous research have been interested in studying graphene's toxicity and biocompatibility both in vitro and in vivo [157]. It has been observed that graphene could induce varying levels of toxicity in animals or cellular models contingent upon the administration routes employed, as well as its traversal across physiological barriers, subsequently leading to tissue distribution or cellular localization [158, 159]. This phenomenon is subject to regulation by an array of factors encompassing lateral dimensions, surface configuration, functionalization, charge, impurities, aggregations, and the corona effect, among others. Several fundamental mechanisms elucidating graphene toxicity have been unveiled, such as physical degradation, oxidative stress, DNA impairment, inflammatory reactions, apoptosis, autophagy, and necrosis [160]

### 6.6.1 Pathways of GO Entry into the Body and Biological Barriers

There are many natural pathways of entry nanoparticles such as ingestion, inhalation and dermal. As a result. We can administer GO by intraperitoneal, intravenous, and subcutaneous injections, as shown in **Fig. 13**.

Also these pathways are used in biomedical applications [161]. GO lead to increasing the rate of respiration in mitochondrial and activating inflammatory and apoptotic pathways , generation of ROS , [161]. For respiratory trac after incoming to the body by intravenous injection, it could be retained in the lung and encourage the construction of pulmonary edema and granulomas [162]. Graphene nanosheets also can destroy the biophysical properties and ultrastructure of surface-active agent film, that is the first step of protection in host, and they show their potential toxicity [163].

macrophages can take nanoparticles dropped at the bottom of the pulmonary alveoli [164]. Rejected by respiratory mucus via the hair cells action [165]. For smallest one of them, go through the pulmonary epithelium and finish in the interstitial liquid [166]. Graphene is also respected to be a very good drug delivery system [167]. Usually insert including anticancer drugs to enhance oral bioavailability [168]. Research shows that the length of intestinal villi in offspring mice significantly affects the concentration of oral graphene (GO). It's recommended to restrict absorption of nanoparticles after oral supervision [169-171]. GO-based nanoparticles are considered antimicrobial agents in ointments and dressings, and graphene aggregates near injection sites [161, 172]. In the toxicity of dermal of graphene, the toxicity is negligeable, so it is necessary to understand the toxicological mechanisms well. Studies have shown that graphene particles which have a diameter (54:9 ± 23:1 nm) have difficulty penetrating Hemat epididymal and the hematotesticular barriers after the intra-abdominal injection. The quality of sperm of the mice wasn't affected, however at 300 mg/kg dose [173]. For the placental mammal barrier, one of the studies recommended that the placenta does not make a block against the incoming NPs to the fetus, exactly alongside the supply of nanoparticles regarding the fetus [174]. On the other hand, the barrier of blood-brain was found to reduce graphene oxide had diameter (342 ± 23:5 nm) is effective. over time, of inserting itself into the inter endothelial cleft and decreasing the paracellular seal in the barrier [175].

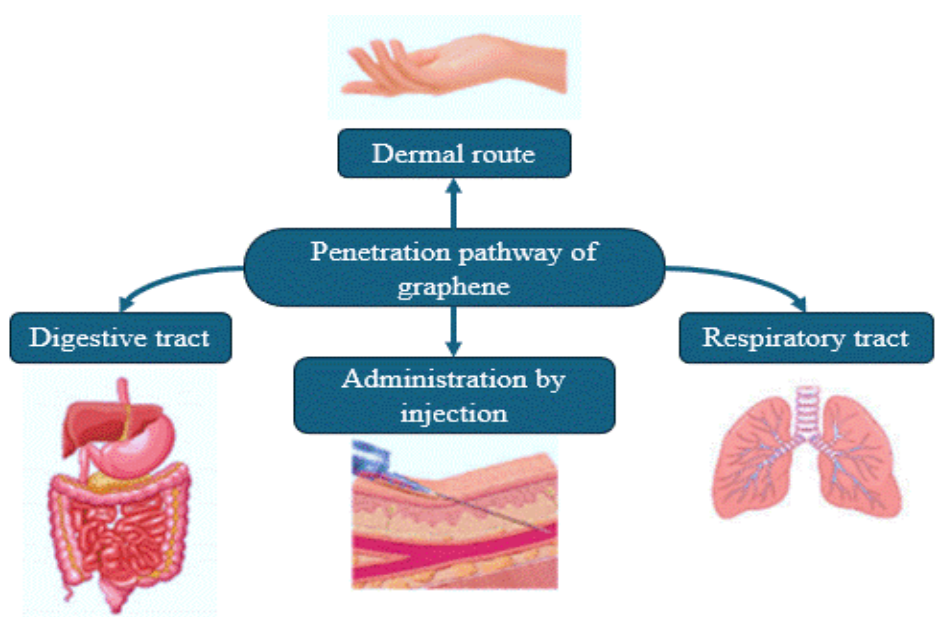

**Fig. 13**. ways of GO pass into the body.

**6.6.2   Possible toxicity mechanisms of graphene**

Although the physicochemical characteristics and toxic nature of graphene have been extensively researched, the precise mechanism behind its toxicity remains vague. The main mechanisms of graphene cytotoxicity are depicted in **Fig. 14**. Within those mechanisms, transforming growth factor-β and toll-like receptors. Tumor necrosis factor-alpha also participates in the signaling pathways network**.** Additionally, oxidative stress is a critical factor in these pathways [160].

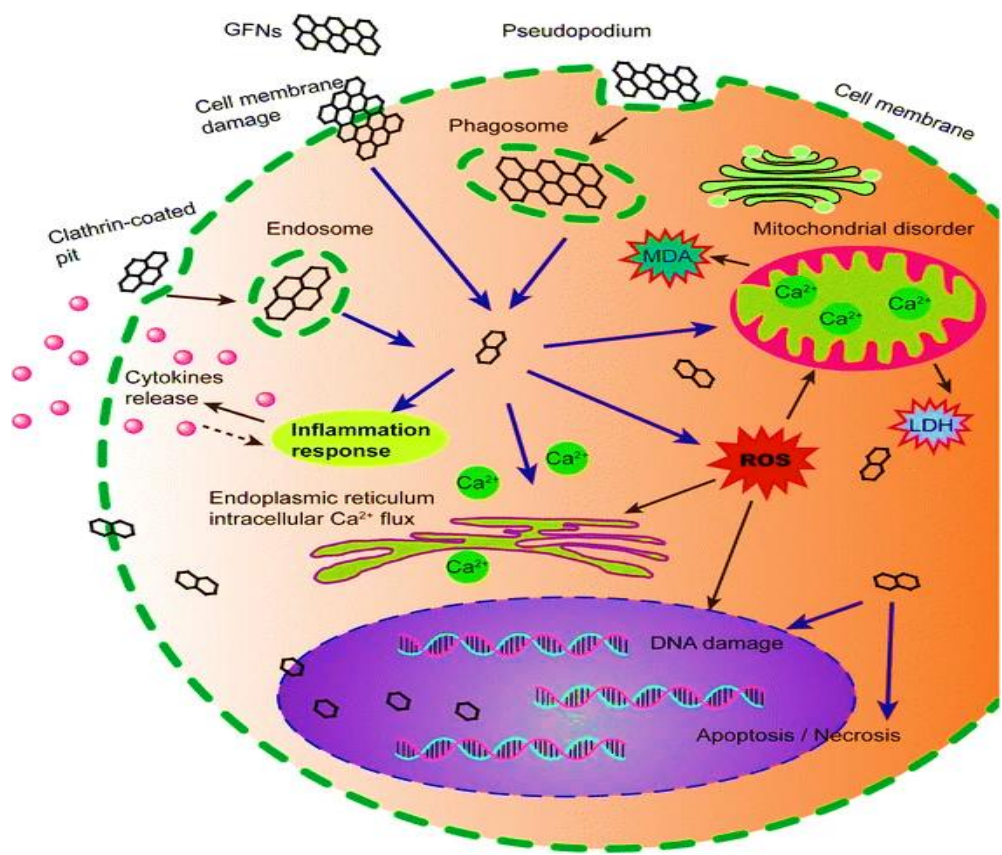

**Fig. 14**. The schematic illustration depicts potential pathways through which graphene may exert toxicity. Graphene penetrates cell membrane, resulting in producing ROS, then increasing LDH and MDA levels, and releasing of $Ca^{2+}$. Subsequently, it causes various forms of cellular damage such as failure of the cell membrane and performance of inflammatory responses. It can also cause DNA fragmentation, distraction of function of mitochondria and apoptosis, or necrosis [160].

**6.6.3   Interactions of graphene with Cell Membranes**

Research indicates that the interaction between graphene and cell membranes is a significant factor in cytotoxicity, with direct contact with E. coli causing damage to the outer membrane, releasing intracellular components, and causing cell death. [176, 177]. Another research study has indicated that graphene's cytotoxic effects stem from its direct interactions with cell membranes, particularly graphene nanosheets, resulting in physical harm to the membranes. Furthermore, exposure of graphene to bovine fetal serum (FBS) has been found to mitigate observed damage, attributed to graphene's exceptionally high protein adsorption capability [177, 178]. Moreover, the interaction of graphene with lipid membranes has been identified as the mechanism underlying the destructive extraction of membrane lipids. Upon penetrating the cell, graphene has the capacity to degrade substantial amounts of lipid membrane phospholipids, thereby inducing cell membrane degradation [179].

**6.6.4 Oxidative Stress:**

Graphene toxicity often results in reactive oxygen species (ROS), leading to oxidative stress characterized by free radicals and antioxidants. These secondary messengers damage cellular molecules, causing DNA fragmentation, protein denaturation, and disruption of mitochondrial functions, as well as lipid membrane damage [180]. Exposure of Euglena gracilis to graphene for a period of ten days led to growth inhibition, reduction in photosynthetic pigments, and increasing of ROS levels [181]. The study reveals that graphene exposure leads to oxidative stress, apoptosis, and DNA injury in human lung fibroblast (HLF) cells [182]. The accumulation of graphene blocks ion channels, resulting in ROS production. Furthermore, treating HL-7702 cells with graphene caused harm to the cell membrane in a manner depending on the dose, as set by LDH release [183]. Small graphene particles can be degraded by lysosomes and excreted without cytotoxic effects. Larger graphene oxides can damage cell membranes by binding to proteins and interacting with phosphatidylcholine, causing ROS generation. Excessive exposure to graphene leads to a decrease in superoxide dismutase and glutathione activity, reducing their ability to prevent ROS accumulation. ROS production in graphene-treated cells activates TGF-beta and MAPK signaling pathways, leading to the activation of apoptosis-inducing members Bim and Bax of the Bcl-2 protein family. This activates caspase 3, leading to mitochondrial fragmentation, DNA damage, inflammatory responses, and apoptosis or necrosis [184].

**7 Fabrication Method**

Graphene, a monolayer of hexagonal carbon atoms, has been studied widely since its breakthrough in 2004 by Jim and colleagues, who magnificently exfoliated single-layer graphene by a repeated exfoliation process [4, 185]. Graphene, a 2D nanomaterial composed of sp² hybridized carbon atoms arranged in a honeycomb lattice, has garnered immense attention due to its exceptional properties. The isolation of graphene has spurred numerous research efforts aimed at synthesizing this material efficiently and at scale [1]. graphene material have been successfully and developed applied to energy technologies like as batteries and supercapacitors [186, 187]. In order to continue graphene research, make the most approaches, lop it, and produce a material with high quality and efficiency, three sections must be studied: top-down approaches , bottom-up approaches and Green synthesis [188, 189].

### 7.1 Top-Down Approaches

These methods start with bulk graphite and employ various physical or chemical processes to isolate single layers of graphene, , as shown in **Fig. 15** [190 ,185]. These top-down approaches are particularly appealing due to their potential scalability and the ability to produce graphene in significant quantities. However, challenges such as defect control, layer uniformity, and maintaining the intrinsic properties of graphene remain areas of active research and development [1], as shown in **Table. 3**.

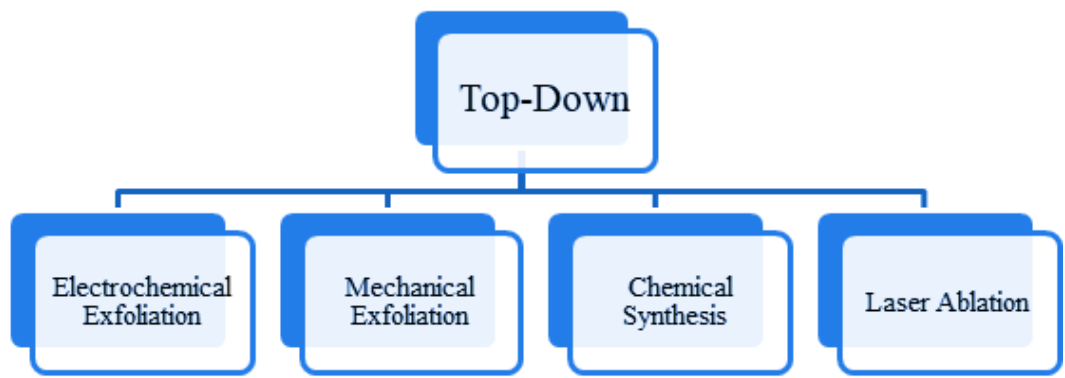

Fig. 15. Examples of Top-Down Approaches

### 7.1.1 Electrochemical exfoliation:

The production of nearly pure graphene with a low O concentration, few layers, and a large wing size is one of the main goals of graphene synthesis. On the other hand, O is frequently introduced into the carbon lattice of graphene during graphite exfoliation in liquid electrolytes at ambient circumstances, causing the graphene to break into minute fragments. The structural characteristics of graphene materials have shown some degree of controllability through the use of electrochemical exfoliation techniques.[191] There are four primary components used in the electrochemical method: the anode, cathode, electrolyte, and power supply. Because it can impact the amount of graphene produced, the selection of cathode, anode, and electrolyte solution is critical to the electrochemical process.[192]

The graphite utilized as an anode and cathode in Liu et al.'s [193] work was obtained from a pencil , as shown in **Fig. 16 (b).** With an applied voltage ranging from +7 V to −7 V, the cathode and anode are submerged in 1 M $H_3PO_4$. The resultant product exhibits nonuniformity in terms of thickness and size distribution. In 2013, Parvez et al.[194] employed graphite as the anode and platinum (Pt) as the cathode. For ten minutes, the electrodes were immersed in sulfuric acid (electrolyte solution) with a potential of +10 V. This technique yielded a product with 1-3 layers and a yield of roughly 60%. [192]

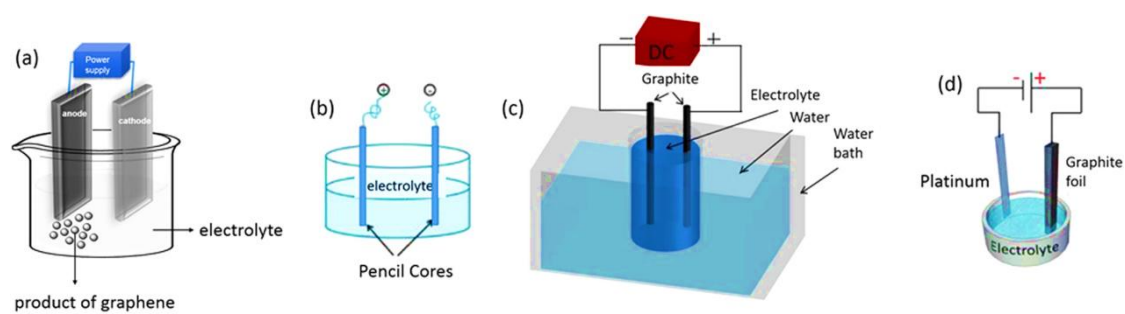

**Fig. 16**. To generate graphene [194], an electrochemical experiment is commonly conducted as follows: (a) based on Liu et al. [3]; (b) based on Hossain and Wang,[5]; and (c) based on Ambrosi and Pumera,[6].

As illustrated in **Fig. 17**, Parvez et al. [195] presented a broad idea for the electrochemical peeling of graphite from inorganic salts. solution [voltage was set to +10 V and the electrodes of graphite were submerged in solution of ammonium sulfate [$(NH_4)_2SO_4$, $H_2O$], as shown in **Fig. 17 (a).** Water reduction at this potential results in the production of hydroxyl ($OH^-$) ions, which are effective nucleophiles. The nucleophiles target the margins and edges of the graphite grains, causing sulfate ions to intercalate into the graphite layer and oxidation processes, as shown in **Fig. 17 (b).** The graphite layer depolarizes and expands due to oxidation processes as shown in **Fig. 17 (c)**. The water molecules co-intercalate with the $SO_4^{2-}$ ions during this reaction. Gas species ($SO_2$, $O_2$, etc.) are produced from the reduction of $SO_4$ ions and oxidized water molecules. The energy of gas species is sufficient to split the layer of graphite and create layered graphene, as shown in **Fig. 17**. The process of graphite exfoliation is also influenced by the electrolyte concentration at the applied voltage. According to Parvez et al., [194] less than 0.01 M of electrolyte ($(NH_4)_2SO_2$) was required for the graphite intercalation procedure to yield 5 weight percent graphene. Concentration increases between 0.01 and 1.0 M will result in graphene products accounting for more than 75% of the total weight. This phenomenon is consistent with the general mechanism of graphite's electrochemical exfoliation in inorganic salts.

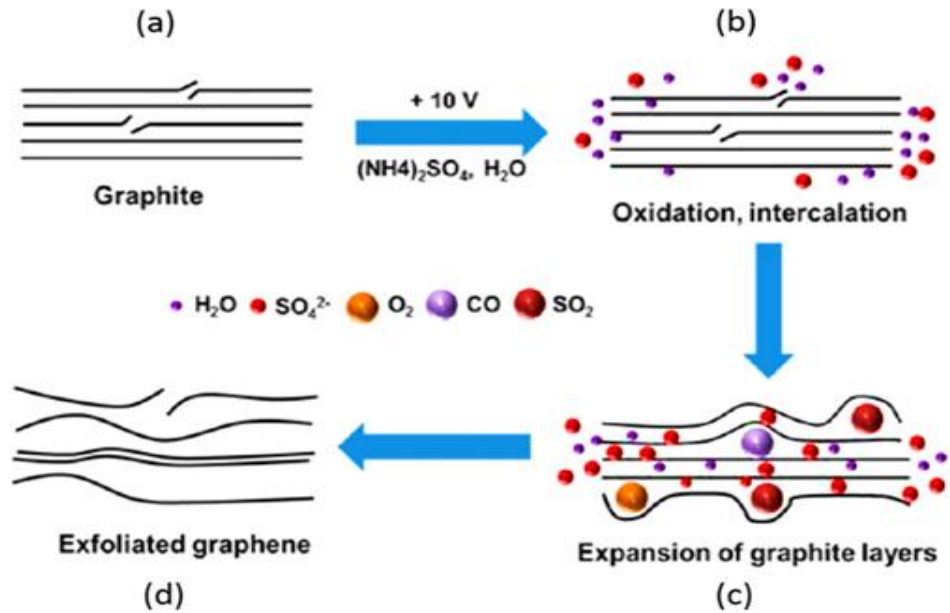

**Fig. 17**. Parvez et al. hypothesized the electrochemical exfoliation mechanism. Graphite is used as the starting material for the following processes: (a) graphene synthesis, (b) oxidation and intercalation, (c) growth of graphite layers, (d) production of exfoliated graphene [196].

Recently, it has been shown that one efficient method for producing graphene is through the electrochemical exfoliation of different carbon precursors, such as graphite rods, carbon sheets, or HOPG. A one-step electrochemical method has been used to manufacture ionic liquid functionalized graphene sheets using graphite rods as electrodes and an ionic liquid/water solution as electrolyte [196]. The electrochemical exfoliation of graphite in ionic liquid solution was studied in detail by Lu et al. [197] and their mechanism was proposed. The process is as follows: (1) water is anodically oxidized; (2) graphite edge planes are hydroxylated or oxidized; (3) ionic liquid anions intercalate between graphite layers to form graphite-intercalation-compounds (GICs); and (4) the GICs undergo oxidative cleavage and precipitate.

Multiple electrochemical exfoliations of graphite rods have been used by Liu J. [198] to demonstrate the improved synthesis of high-quality graphene in a vertical cell design. The process of multiple electrochemical exfoliations, or MEE, was used to create graphene flakes. In a nutshell, the positive electrode and carbon supply were made of graphite rod from used Zinc Carbon dry cells. Protonic acid (such as $H_3PO_4$, $H_2SO_4$, or $H_2C_2O_4$) aqueous solution was used as the electrolyte in the electrochemical cell, with graphite rod serving as the anode and platinum wire serving as the cathode. A certain voltage was applied across the electrodes to start the electrochemical exfoliation process. The ultrasonicated, cleaned, dried, and gathered water-soluble graphene flakes. Additionally, typical (parallel) electrochemical arrangements were used for control studies.

Bakhshandeh and Shafiekhani [199] documented how ultrasonic waves affected the characteristics of graphene that was produced electrochemically. It was discovered that the generation of graphene was directly impacted by ultrasonic waves. Electrolyte solutions can be homogenized by ultrasonic waves, and graphene structures can have their oxygen groups reduced. The decrease in graphene structural flaws is directly correlated with the ratio of reduced oxygen groups.

Hossain and Wang [200] also looked at the impact of temperature (range from 25°C–95 °C) when using the electrochemical-sonication method to produce graphene. Raman spectroscopy characterization reveals that as temperature rises, graphene products exhibit the tiniest flaws. The results show that throughout the synthesis process using the combination electrochemical-sonication approach, researchers and companies use elevated temperature and controlled ultrasonic waves to achieve the minimum flaws.

According to Lee J-W et al.[201] a highly effective spontaneous reaction procedure devoid of oxidation and reaction was able to manufacture a high-quality GNS with a conversion rate of 95% (100% of theoretical). The production of graphene nanosheets (GNS) involves two spontaneous sequential reactions: the Lithium reaction, which produces Li intercalated graphite intercalation compounds (GIC) through direct contact with Li flakes and graphite powder, and the Li intercalated GIC reaction with water, which produces GNS. Although the electrochemical lithification process does not require electricity, it is quick and efficient, making it a crucial component in the production process.

As shown in **Fig. 18**, P.C. Shi et al. [201] have developed a non-electrified electrochemical method to generate high-quality graphene layers on a large scale in a 1 M LiPF6/propylene carbonate (PC) electrolyte. PC-based electrolytes can efficiently and productively exfoliate HOPG into a few graphene layers. Every time a Li∥graphite cell is discharged, Li+ -(PC)4 dissolves in the graphite layers through electrochemical intercalation. This process generates multiple Li∥graphite microcells and Li+ -(PC)4 solutes, resulting in NEEG, without consuming electrical energy. The easy random assembly of Li||graphite microcells allow for continuous graphite exfoliation, resulting in high yield of graphene fabrication. Reactivity technology uses graphite powders instead of a graphite electrode, making it beneficial for large-scale graphene synthesis.

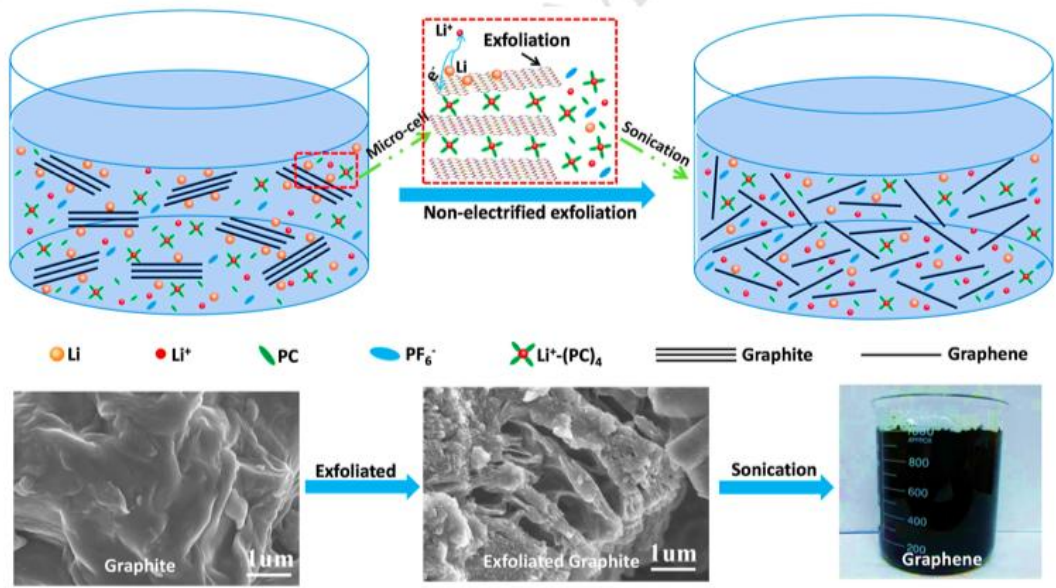

**Fig. 18**. Schematic illustration of the non-electrified electrochemical exfoliation method for large-scale production of high-quality graphene sheets in 1 M LiPF6/PC electrolyte [201].

**7.1.2  Laser ablation:**

A novel one-step technique for creating nanomaterial, and particularly graphene, is laser ablation. This approach has several encouraging benefits, such as being safe for the environment, having simple experimental conditions that do not call for harsh conditions, producing nanoparticle products free of undesired contaminants, and not requiring the use of hazardous synthesis reactants, as shown in **Fig. 19** [192].

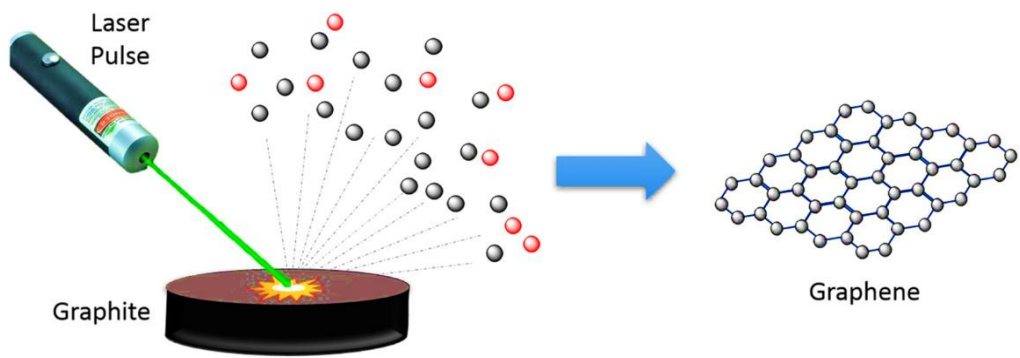

**Fig. 19**. Mechanism of graphene synthesis by laser ablation [192].

In two stages, normal pressure and at room temperature, R. Hameed et al. [202] have successfully prepared graphene sheets. The first step involved laser ablation of a pure graphite pellet in a liquid process, and the second involved re-irradiating suspensions of carbon nanotubes obtained with the same laser energy after the target was removed.

M. Jalili et al. [203] developed an environmentally friendly, one-pot method for graphene synthesis using nanosecond laser ablation of three targets at different wavelengths in liquid media. The study identifies optimal parameters for high-quality few-layer graphene nanosheets. Laser exfoliation of FG1 and FG2 produces micrometer-sized nanosheets, which are chosen for laser ablation. The higher penetration depth increases an evaluation efficiency at 1064 nm but results in unwanted nanoparticles. The optimal wavelength is 532 nm. Among the three solvents, few-layer graphene created by laser ablation in acetone medium is chosen due to its non-aggregation.

The liquid-phase LA experimental setup used by Rosemary L. Calabro et al. [204] to create GQDs is shown in **Fig. 20**. nCNOs were typically crushed into a pellet at 15,000 psi and maintained under pressure in an experiment. Because it allowed the pellet to stay intact throughout the entire ablation process, this pressure was used. After moving the particle into a vial, 3 mL of deionized water was added. To reduce evaporative loss of the deionized water, a quartz lid was utilized. 532 nm light was produced using a Spectra-Physics Quanta-Ray® GCR-3 Q-switched Nd:YAG laser fitted with a harmonic generator.

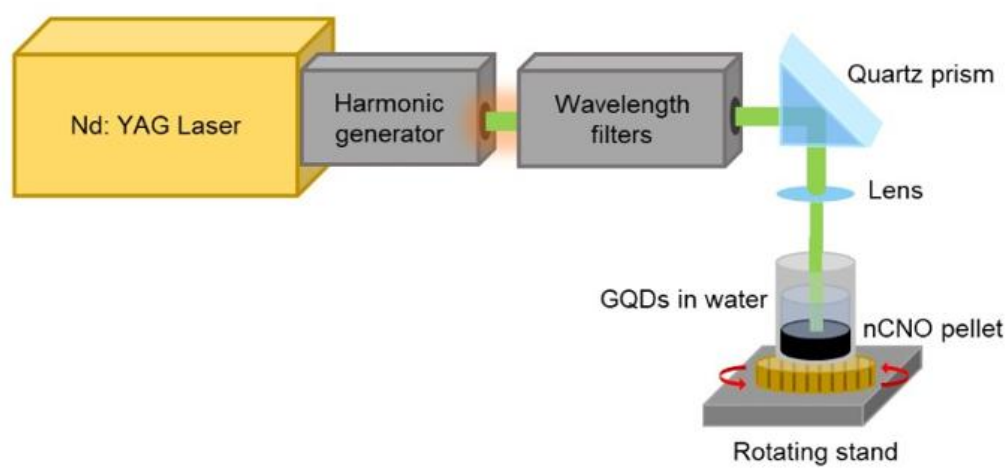

**Fig. 20**. Schematic illustration of a laser ablation setup [204].

M.M. Juvaid et al. [205] use an excimer laser ablation of graphite in the DI water medium as part of a novel single-stage synthesis method to realize bilayer graphene. By substituting commercially available graphite for the HOPG target, this technique is less expensive and more practical for device applications. We discovered that whereas higher laser fluence (> 2.5 J/cm$^2$) permits the creation of graphitic films, lower laser fluence favors the synthesis of bilayer graphene. Furthermore, by adjusting the laser's fluence, repetition rate, and exposure time, one may vary the size and layers of produced graphene, as shown in **Fig. 21**.

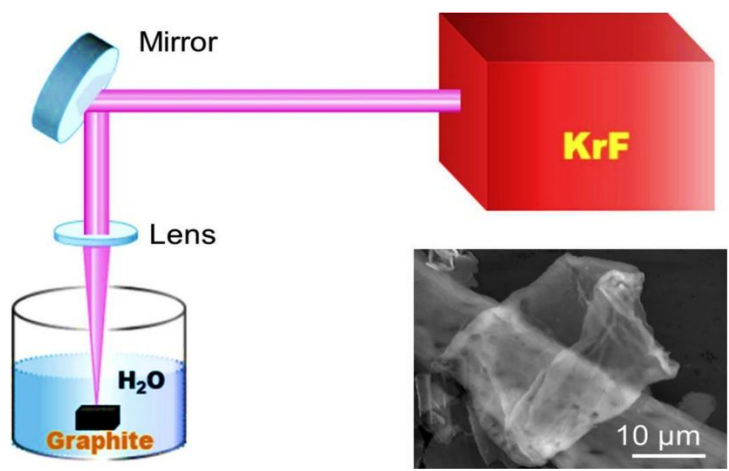

**Fig. 21**. Schematic illustration of irradiation technique based on KrF excimer laser ablation of graphite in DI water along with the SEM image of ablated graphene sheets [205].

Avramova, I. et al. [206] studied the PLAL technique using a 6% H2O solution as a model system. They demonstrated that this approach could produce irradiation solutions containing nanoscale carbon phases, such as graphene and graphene-like (rGO, GO, and defective graphene) and clusters of graphene-like flakes. The unique approach of PLAL, which takes place inside a quartz tube isolated from the outside world, allows for the conversion of treated carbon materials with different atom types. The broad use of this approach is made possible by the controllability of the medium, laser wavelength, flow velocity, and irradiation energy flux. The suggested approach has several advantages, including the ability to efficiently functionalize vast volumes of nano-sized carbon phases and being simple, easy to apply, and simple. It has been determined that established flow-mode PLAL procedures are reproducible.

Using readily accessible and reasonably priced graphite flakes as the carbon precursor, S. Kang et al. [207] effectively demonstrated that N-doped GQDs with outstanding optical properties (e.g., QY, PL intensity) can be produced by the PLAL process. When the traditional PLAL method (without sonication) is used, the nitrogen atoms may be doped into the GQDs more readily, leading to the significantly increased QY of the off-NGQDs (0.8%: 9.1%) after doping with N.

By using laser ablation of the graphite target in a single step, F. Kazemizadeh et al. [208] created porous graphene. As shown on **Fig. 22**, depicts an experimental setup schematic. Using ball milling method, graphite and nickel particles were mixed to form powder at a concentration of 2%. Powdered nickel was employed as a catalyst-assisted material to form a sp$^2$ nanostructure. PGs were created using an ultra-high flow rate of 2.4 l/min of argon gas, which is nearly ten times the amount typically applied to the synthesis of CNTs.

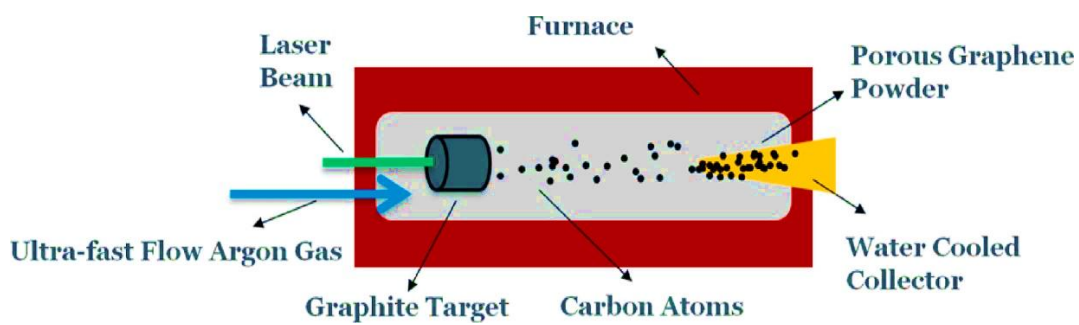

**Fig. 22**. Schematic illustration of the experimental setup [208].

**Table. 3** Some technique type (Top-Down Approaches).

| Technique type | Synthesis method | Product | Properties | Applications | Refs. |
|---|---|---|---|---|---|
| **Top-Down Approaches** | Mechanical exfoliation | Monocrystalline graphitic films | Direct, high structural and electronic quality, simple, low cost | FET devices | [4, 209, 210] |
| | Chemical exfoliation | Few-layer graphene | Direct, low-cost, simple large-scale production, high yield, practicability of sample handling (liquid suspension) | sensing device | [209-212] |
| | Electrochemical exfoliation | graphene sheets. | Lower mass production costs does not require harsh chemicals, | energy storage devices | [213] |
| | Liquid-Phase Exfoliation | graphene sheet (monolayer or FLG) | Control the size, thickness distribution and good quality at a large scale | Solar cells | [214] |
| | Electrophoretic deposition | Graphene nanosheets | Bending strength 477 MPa and tensile strength 518 MPa | Supercapacitor | [1, 215] |
| | arc discharge method | Graphene sheet | a product with good thermal stability and high electrical conductivity | energy storage devices | [188, 216] |
| | Direct sonication | Both single and multiple layers | Low-cost and unmodified graphene | Efficient and precise sensor | [189] |
| | Mild chemical reduction of GO aqueous | graphene aerogels | Inexpensive and mechanical stable, density range from 3 to 5 mg/cm3 and a porosity as high as 99.8% | Refrigeration devices | [187, 217] |

### 7.2 Bottom-Up Approaches

It refer to the synthesis methods that build graphene from smaller molecular precursors. These methods involve epitaxial growth, chemical vapor deposition (CVD) and pyrolysis , as shown in **Fig. 23** and **Table. 4**. It have emerged as promising methods for producing high-quality graphene with controlled properties by manipulating the synthesis conditions and the molecular building blocks used [190, 218]. The drawbacks for bottom-up methods include lower yield, higher cost and the difficulty to scale up compared to top-down methods [188].

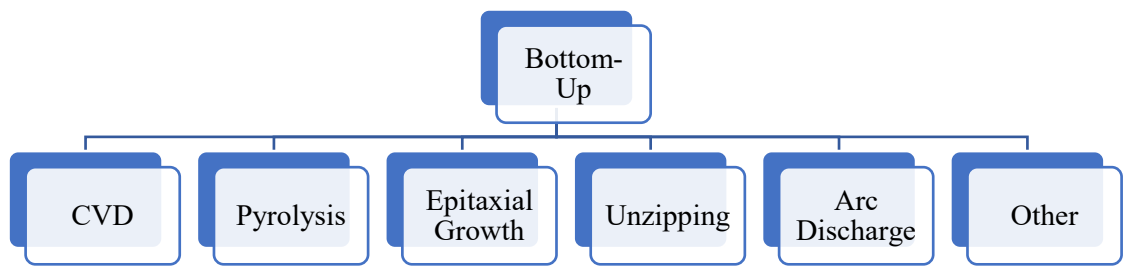

**Fig. 23**. Examples of Bottom-Up Approaches.

### 7.2.1 Chemical Vapor Deposition (CVD)

CVD is a commonly used technique for producing high-quality graphene, which is crucial for optical and electronic applications [219-222]. CVD involves growing a molecule of gas on a substrate under specific environments, such as pressure, temperature, and rate of gas flow [221, 223-227]. Two main approaches are plasma-enhanced CVD (PECVD) and thermal CVD [228]. Thermal CVD uses a vacuum pump, vacuum tube, mass flow controller, vacuum gauge and high-temperature furnace, while PECVD uses plasma to disintegrate the gas source and react on the surface of substrate, producing sheets of gr Naphene with low temperatures and pressure [229-232]. The deposition process occurs on thin films, crystalline, solid, liquid, or gaseous precursors, typically on metal substrates like copper, nickel, Pd, Ru, and Ir [233-240]. CVD techniques have led to innovations in energy-related materials and information [241].

However, Dry and wet chemical approaches can be used to overcome challenges including additional contamination throughout transfers, destroying sheets of graphene and challenges with reusing metal substrates. Thus, CVD is the best technique to produce single crystal graphene that has fewer carbon impurities, and the quality of the products is high [240, 242-245].

Da-Je Hsu et al. [246] prepared a graphene nano walls (GNWs) for high-voltage supercapacitors by PECVD reactor powered by a power supply and a magnetron operating at 2.45 GHz microwave frequency. The power of microwaves ranges from 250 to 1400 W. To create the plasma torch, the magnetron used a waveguide to activate the complex gas of $CH_4$ and Ar in the quartz tube. The inner diameter of the quartz tube is 25 mm, and the outer is 28 mm. The $CH_4$ flow rate is 20 to 80 sccm, and Ar flow rate is 20 to 60 sccm.

V. Thirumal et al. [247] developed multilayered graphene nanofibers (GNFs) for supercapacitor devices by preparing Ni foil and SS foil, washing and sonicating them with water, ethanol, and acetone. Ni-NiNO$_3$ was synthesized by drop casting coatings of NiNO$_3$ (5M) on Ni foil and drying them at 100°C for 4 hours. The substrates were placed in a CVD tube, as shown on **Fig. 24**, cleaned, and synthesized using argon, hydrogen at room temperature and acetylene hydrocarbon source at 700°C, and cooling the CVD with argon flow.

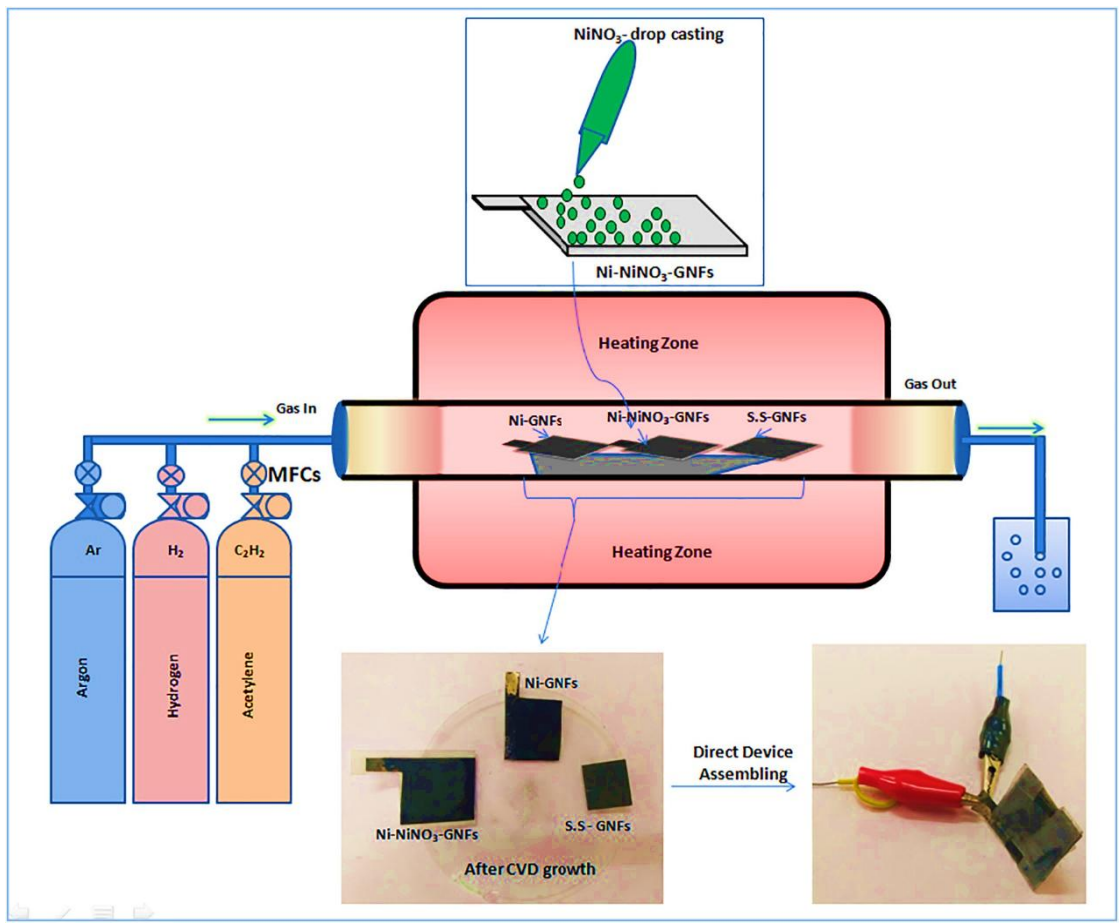

**Fig. 24**. Schematic diagram of CVD technique for graphene nanofibers growth by drop casting coating of NiNO$_3$ on Ni-foil for supercapacitor devices [247].

Andrea Capasso et al. [248] synthesized graphene films on Cu substrate by CVD technique with $CH_4$ as a source of carbon. The Cu substrate was put into the CVD and then heated to 1060°C with flowing of argon for 10 min to annealing process. The graphene was synthesized at 1060°C for 10 min after annealing by flowing of argon, methane, and hydrogen. Cool the CVD to 120°C before taking out the films to protect the surface of substrate from oxidation.

Elif Ceylan Cengiz et al. [249] demonstrates the creation of multiple layers graphene films on a nickel foil with thickness 50 mm using Chemical Vapor Deposition (CVD). At 1000°C, the film started annealing under Ar, $CH_4$ and $H_2$ gases to remove the oxide layer. The specimens were cooled to room temperature using flowing of argon gas after a 5-minute growth period. The "fishing technique" was then used to create multiple layers of graphene on a glass fiber separator. The foil was submerged in a dense $FeCl_3$ solution, then put in DI water. The paper was then submerged and dried at 70°C for 4 hours.

Jianhang Yang et al. [250] created graphene foam for lithium-ion batteries by cutting Ni foam into parts, washing it with ethanol, drying it, and placing it in a CVD tube with a mixture of hydrogen and argon. The process started at 100-200°C, then Ethylene was added at 1100°C. After cooling to 200°C, samples were removed from the CVD, as shown on **Fig. 25**, and Ni was removed by immersing the sample in a $FeCl_3$ solution to obtain pure graphene. The graphene foam was then cleaned multiple times with DI water and ethanol.

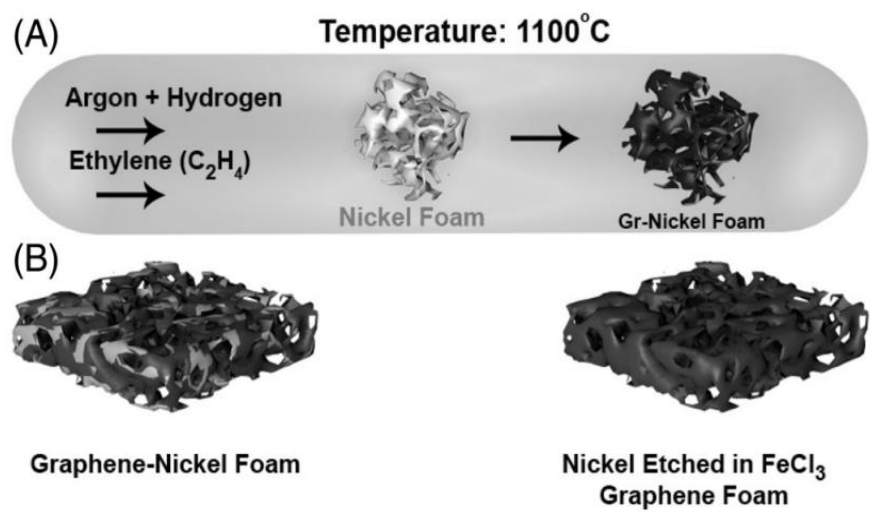

**Fig. 25**. (A) Growth of graphene on Ni foam in CVD at 1100°C. (B) Graphene foam after Ni etched in $FeCl_3$ [250].

Gyujeong Jeong et al. [251] use the electron beam evaporation method to deposition of Cu on a photoresist patterned substrate. The Cu grid pattern was fabricated when removing the photoresist. Graphene was fabricated on Cu foil through CVD technique. Cu foil was put into CVD and reached the temperature to 1000°C for 30 min with flowing of $H_2$ to annealing. Then, graphene was grown by the flowing of methane.

Yifei Ma et al. [252] demonstrates the synthesis of vertically oriented graphene (VG) using a quartz plate in a tube-type inductive coupled PECVD system. The process involves cleaning the chamber with hydrogen gas, plasma generation, and growth of VG under various conditions. VG-based supercapacitors were created using polyvinyl alcohol (PVA) as a polymer electrolyte. PVA powder has been dissolved in water, heated, and added to a solution. The PVA/VG sheet was then spin-coated onto the VG surface, forming a VG/PVA/PVA/VG configuration. Cu surfaces were used as current collectors to pack the stacked layers.

Neeraj Mishra et al. [253] synthesized graphene on both $H_2$-etched and pristine sapphire where pristine sapphire was cleaned and $H_2$ etching occurs in growth reactor with flowing of Ar. Annealing of substrate at 1200°C for 10 min with flowing of Ar, then $H_2$ and $CH_4$. Finally, cool with Ar flow, as shown on **Fig. 26**.

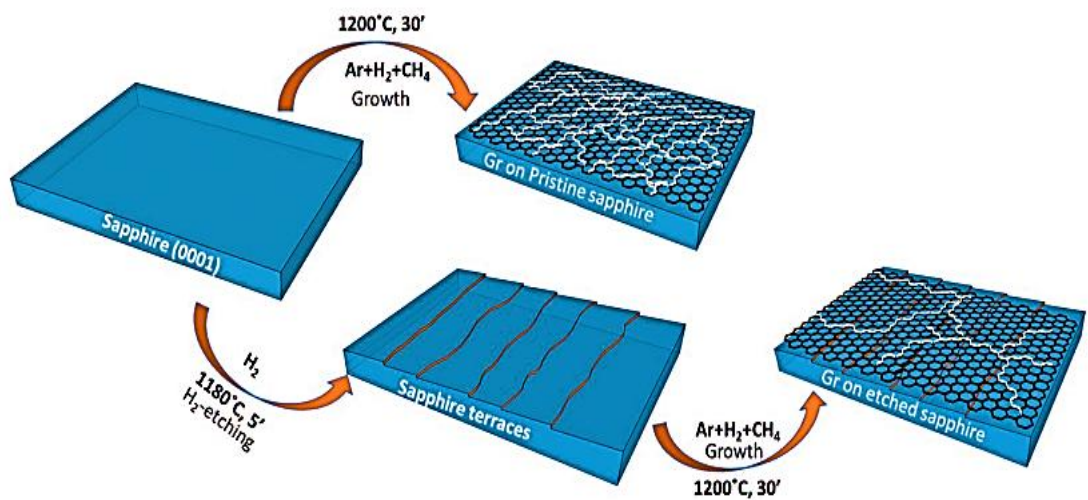

**Fig. 26**. Growth of graphene on both $H_2$-etched and pristine sapphire [253].

Lixuan Tai at al.[254] synthesized graphene on monolayer Si substrates in CVD with vacuum system and Ar flow. And heated the substrate to the required temperature with flowing of Ar and $H_2$. Graphene growth by using $H_2$ and $CH_4$. Finally, cooled the chamber at room temperature.

Xinghai Liu et al.[255] fabricated Cu@GNS/Al by CVD. First, the dry Cu/Al powders were synthesized, transferred into quartz tube and heated at 364°C with $H_2$ flow. And heated to 600°C with flowing of Ar and $CH_4$. The carbon atoms were grown on Cu surface to fabricate graphene nanosheets (GNS). Finally, Cu@GNS/Al was synthesized.

Kailun Xia et al.[256] fabricated a single-layer graphene on Cu film by transferred a Cu film into the furnace and heated to 1070°C with flowing of Ar, $H_2$ and then, $CH_4$. The growth was finished by ceasing $CH_4$ flow and cooling the system.

**7.2.2 Epitaxial Growth**

Epitaxial growth is a method of deposition layers of graphene on a crystalline substrate, such as hexagonal carbide (SiC), boron nitride, Ir, Ru, Cu, and Pt. It offers advantages such as being designed by lithography techniques and having high-quality graphene [257]. The epitaxial film and substrate can be heteroepitaxial or homoepitaxial, based on the material [258]. Atoms of carbon graphitize after Si atoms sublimate from a SiC face in the epitaxial growth of graphene on SiC. Extremely high vacuums and high temperatures are necessary for this kind of growth. [259, 260].

Epitaxial growth on SiC is considered a heteroepitaxial film of monolayer graphene and it is a hopeful technique due to its flexibility, electronic properties, and high-quality graphene [258]. Efforts have been made recently to create extremely pure and smooth graphene films on the upper part of a SiC surface. High temperature in the method of sublimation may cause defects of "terrace steps" on the SiC substrate. To address this issue, polymeric carbon source assisted sublimation growth can be used to stabilize the SiC substrate [261].

Epitaxial graphene is a hopeful technique for wafer-scale integration of graphene-based electronics, providing high crystal quality graphene for various applications such as integrated circuits, transistors, UV photodetectors, biosensors, programmable electrical systems for resistance metrology and gas sensors [262-268].

The epitaxial growth on SiC is depend on the disintegration of the SiC crystal surface, which is affected by sublimation and morphology aspects [268, 269]. Different configurations were obtained when graphene was grown on the basal plane of the SiC (0001) Si face of the hexagonal SiC substrate as well as the SiC (0001) carbon face. Because Carbon faced graphene has better properties [260, 270].

Zezhao He et al. [271] created graphene FETs on Si terminated 4HeSiC (0001) substrate using an improved photoresist free process. They deposited a 20-nm gold metal sheet on the bilayer graphene surface, protecting the graphene surface and avoiding contamination from photoresist. The gadgets' active region was defined using oxygen plasma etch. Liftoff and metal evaporation were used to prepare the source and drain contacts. A 2-nm Aluminum metal sheet was deposited on graphene, followed by a 20-nm $Al_2O_3$ film. The $Al_2O_3$ gate dielectric was improved by quick thermal annealing at 400°C for 2 minutes. The gate electrode was evaporated, and the EBG-FETs were created, as shown on **Fig. 27**.

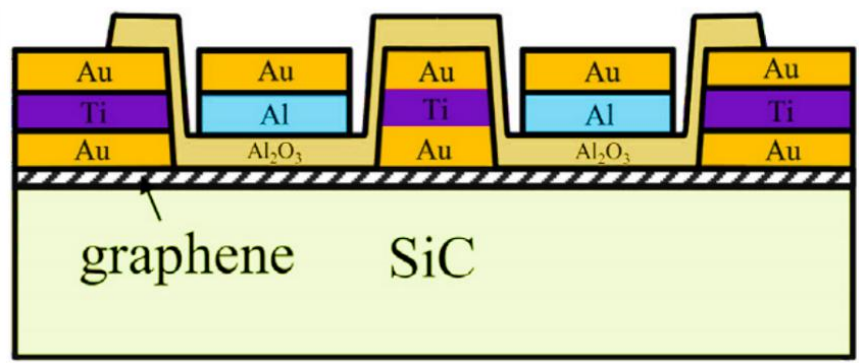

**Fig. 27**. Schematic illustrations of EBG-FET on SiC substrate [271].

Christos Melios et al. [272] created a sensor device using SiC-grown epitaxial graphene on semi-insulating 6H-SiC surfaces with a resistivity >1010 Ω cm$^{-1}$. The surfaces were 8x8 mm2 and misoriented. The graphene was fabricated using Si sublimation with an inert gas overpressure of Ar, etched in $H_2$ at 100 mbar, evacuated, and Ar added. The device was fabricated at 1580 °C for 25 minutes in an Ar atmosphere, then cool down in Ar to 800 °C. Three steps were taken in the fabrication procedure of the gadget, including defining electrical contact pads, electrical leads, and Hall bar design.

Ruizhe Wu et al. [273] synthesized graphene on Cu substrate after washed and dried in $N_2$ gas and transferred into the quartz tube. Growth graphene started at 1040°C with Ar flow. Then, annealing of sample with flowing of $H_2$ and $CH_4$. Finally, the system was switched off and the sample was cooled to Ambient temperature.

Bing Deng et al. [274] synthesized graphene on CuNi by heating the film to 1000°C with flowing of Ar, $H_2$ and $CH_4$. After growth graphene, $CH_4$ flow was stopped. The sample was cooled at room temperature.

Xuefu Zhang et al. [275] fabricated graphene on Cu/Ni (111) Film when film heated to 750°C under flowing of Ar, $H_2$ and then $CH_4$. The sample was cooled at room temperature after CH4 flow stopped.

D. Momeni Pakdehi et al. [276] performed a growth graphene on 6H- and 4H-polytype of SiC substrate. The process of growth was performed in furnace where annealing occurs at 900°C, 1200°C and 1400°C. The graphene grown at 1750°C with Ar flow. The sample was cooled at room temperature.

Stefan A. Pitsch et al. [277] fabricated graphene on 6H-SiC substrate. The samples were subjected to a sublimation process in a furnace with Ar flow, starting at around 1500°C with Ar. The growth process occurs in 2070°C.

Zhizhuang Liua et al. [278] grew graphene on 4H-SiC. The substrate was put in a vacuum system with Ar pressure. The substrate is irradiated using beam of laser to increase the temperature until it reaches between 1500°C and 1780°C to obtain graphene.

Atasi Chatterjee et al. [279] synthesized graphene on three different 6H-SiC substrates. The substrates were inserted into the reactor to growth graphene with vacuum system and Ar flow at 400°C and annealing of samples in 1750°C to obtain high-quality graphene.

**Table. 4** Some types of Bottom-Up Approaches technique.

| Technique type | Synthesis method | Product | Properties | Applications | Refs. |
|---|---|---|---|---|---|
| Bottom-Up Approaches | CVD & epitaxial growth | few-layer graphene films on metal substrates such as Ni and Cu | Large-scale production, high quality, uniform films, tailoring of graphene quality possible | electronic devices | [210, 211, 280] |
| | CVD | few-layer polycrystalline graphene with a larger area | very-high-quality, large area, and single crystal graphene | electronic applications | [213] |
| | Plasma Chemical Vapor Deposition | High quality graphene films on transition metal substrates | Low cost, low temperature required and high optical properties (Transmittance 78–94%) | electronic devices & optical devices | [1] |
| | Metal–organic chemical vapor deposition (MOCVD) | Single layer graphene | Optical transmittance> 97% and sheet resistance of 450 ± 47 X/sq | Transparent conducting electrode | [1, 281] |
| | Thermal Decomposition | graphitic layers | High-area single layer graphene films on SiC at normal atmospheric pressure and 1650°C. | Electronic devices | [1] |
| | Spray pyrolysis | Graphene thin film | thin and homogeneous graphene film can be produced. Additionally, no high temperatures are required as the glass substrate only needs to be heated up to 200°C. | Optical devices | [189, 213] |
| | hydrothermal | Graphene nanoparticles. | Fast process, low cost, good stability, and high quality | Capacitors and batteries | [187] |

## 7.3 Green Synthesis Method

Green synthesis of graphene typically employs plant extracts, biomolecules, and other natural resources as reducing agents to convert graphene oxide (GO) into graphene. This eco-friendly route not only avoids the use of hazardous chemicals but also offers a profitable choice of conventional synthesis methods. The use of ascorbic acid, L-cysteine, and various plant extracts has been reported to yield high-quality graphene while being benign to the environment [284-282]. The integration of green chemistry principles into the synthesis of graphene represents a considerable step forward in the field of nanomaterials. It aligns with global efforts to reduce pollution and conserve resources, making graphene an even more attractive candidate for energy storage, electronics, and biomedicine applications [1, 188].

Centeno et al. [285] synthesized graphene quantum dots GQDTs by using Opuntia sp. extract and ethylene glycol via pyrolysis microwave method. With the use of this environmentally friendly synthesis technique, we can carefully prepare nanostructures that can be utilized to boost fluorescence quenching at pH = 8.8 that have strong dispersion and narrow size distribution, detect phosphates and phytic acid, and more, as shown on **Fig. 28**.

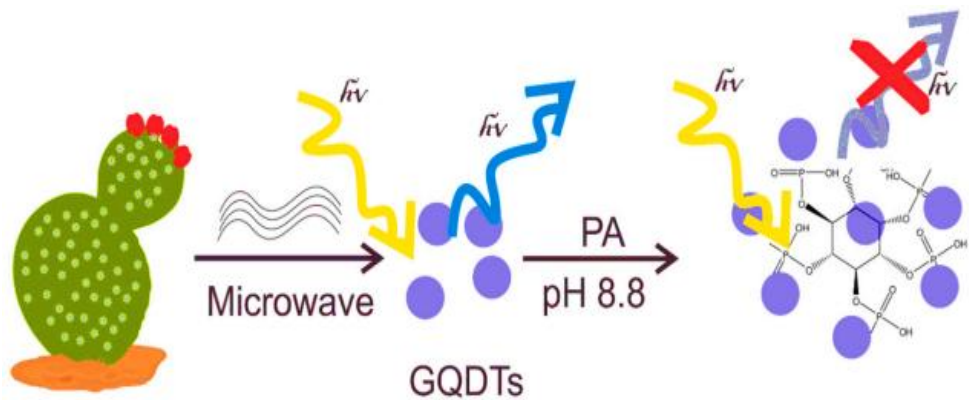

**Fig. 28**. Diagram showing how graphene quantum dots are created and phytic acid is detected [285].

Coros et al. [286] used thermal reduction with different temperatures to prepare graphene oxide GO. There are three steps in the process of obtaining thermally reduced graphene oxide (trGO): Graphite oxide is first produced by oxidizing graphite powder. Second, under sonication, graphite oxide is disseminated in water and exfoliated to form a few layers of graphene oxide. Thirdly, the trGO flakes are made by thermally reducing graphene oxide. At room temperature, the produced materials show significant promise in atenolol adsorption from wastewater.

Tewatia et al.[287] prepared graphene oxide by reducing it with ascorbic acid, also referred to as vitamin C. By increasing the amount of $KMnO_4$, $NaNO_3$ is eliminated in the Modified Hummer's method, which generates graphene oxide. By varying the ascorbic acid content, the degree of reduction can be altered without compromising the properties of reduced graphene oxide.

Gürünlü et al. [288] fabricated graphene by heating graphite in a LiCl-KCl molten salt combination. Graphite was shown to be capable of forming nanostructured carbons in an ionic molten salt (MS) media. By adjusting the synthesis parameters, the technique produces graphene that is purer. By using XRD and Raman investigations, one layer and a few layers were discovered to be the layer counts of graphene products.

Chen et al. [289] utilized cellulose, a naturally occurring polymer, as a novel precursor for the first time to prepare GQDs by using water as well as cellulose only. The produced GQDs strong photoluminescent (PL) qualities, favorable hydrophilicity, little cytotoxicity, and superior biocompatibility have made them useful for bioimaging, as shown on **Fig. 29**.

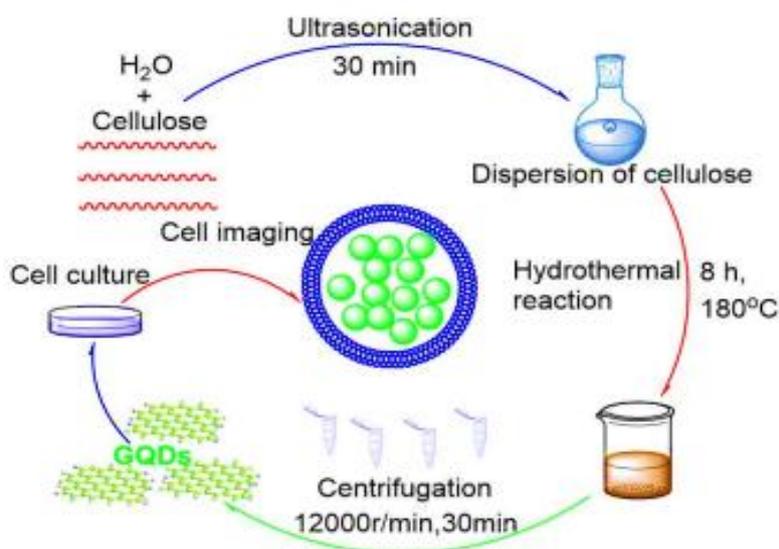

**Fig. 29**. The environmentally friendly path for cellulose-based green synthesis of GQDs [289].

Baweja et al. [290] synthesized Graphene oxide from two varying precursors synthesized for example, graphite flakes GO-1 and sugarcane bagasse GO-2. Two types of graphene quantum dots, called GQD-1 and GQD-2, were created using $KMnO_4/H_2SO_4$ in the oxidized shearing process, with GO-1 and GO-2 as precursors. With a generated yield of 12.54%, the optical characteristics demonstrate the vivid blue luminous nature of graphene quantum dots.

Tai et al [291] prepared GO using Hummers' technique, then prepared green rGO using green tea extract, stressing the objectives of utilizing green tea to synthesis the rGO at various stirring times, analyze its optical characteristics, morphologies, and crystalline structure. Through the display of peaks at 9.75 nm and 25.42 nm, respectively, the XRD spectra have demonstrated the effective synthesis of GO and rGo.

Chamolia et al [292] produced graphene nanosheet GN by reducing graphene oxide using urea, a green reducing agent. Its ability to reduce is analyzed at low temperatures and with varying concentrations. The optical and electrical properties of coated films are also measured subsequent to the creation of transparent conducting films (TCFs) with synthetic GNs.

Sadak et al [293] prepared graphene paper GrP that is flexible, free-standing, and extremely conductive using an environmentally friendly process. They also showed how to use it as an oil and organic solvent absorbent and in supercapacitors. GrP is produced by electrochemically exfoliating graphite to create partially oxidized graphene (po-Gr), as shown on **Fig. 30**.

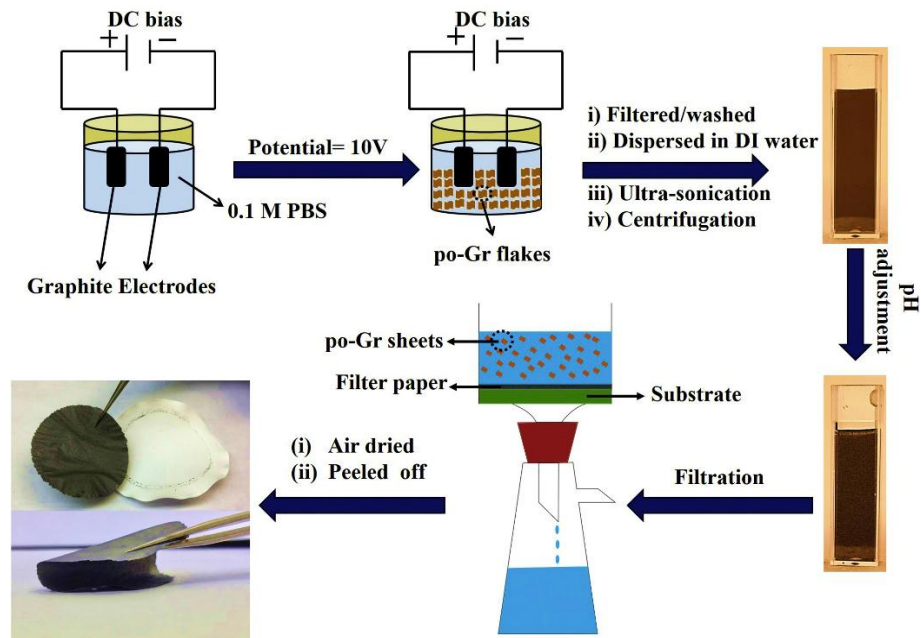

**Fig. 30**. Procedures for creating flexible, free-standing graphite sheets initially [293].

Gan et al. [294] employed an easy-to-use and effective process for the green reduction of GO to reduced graphene oxide (rGO), which has been distributed on bagasse from sugarcane and produced by rGO/bagasse. Furthermore, rGO prepared from bagasse was thought to be a potential methyl blue adsorbent, and the creation of rGO/bagasse material offered significant promise for the removal of dye waste.

Alshamsi et al. [295] used a non-toxic, commercially accessible extract from the aqueous solution of Hibiscus SabdarriffaL as a new model of reductant, produced reduced graphene oxide. Excellent thermal properties and high suspension stability were demonstrated by the RGO with a few-layer structure, making it a potential material for use in industry. Measuring the probability of decreasing functional aggregates in graphene oxide with a natural agent instead of hydrazine is safer, less volatile, and less toxic than with other agents.

Halder et al [296] investigated an innovative one-pot green synthesis method that didn't use any harsh reagents to produce luminous graphene-based quantum dots (GQDs). Using hydrogen peroxide as a catalyst, graphene oxide is hydrothermally synthesized in two hours without the need for additional post-purification procedures. The as-synthesised GQDs exhibit superior biocompatibility and great photostability, as demonstrated by cell survival tests conducted on three distinct cell lines: macrophages, endothelial cells, and a model cancer cell line. It is extremely beneficial for bio-imaging and other associated uses like medication delivery and diagnostics.

S. Ahmed et al. [297] fabricated low-cost and green graphene quantum dots GQDs by using corn powder which dispersed in ethanol, then heated at 200 °C. The application of GQDs as an electron transport layer, in addition to passivating the oxygen agents on the ZnO surface, green produced GQDs quickly remove electrons from the PVK layer to lessen the recombining of electron and hole pairs.

Mahiuddin and Ochiai [298] reduced graphene oxide by using lemon juice as a reducing agent for graphene oxide. The conjugated systems were regenerated, and the oxygen-producing functions of GO were efficiently removed by the reduction with lemon juice, as demonstrated by X-ray diffraction and UV-vis and FTIR spectroscopy studies. Utilizing methylene blue (MB) as a model, the application of the resultant rGO as an organic contaminant's adsorbent was investigated.

Haydari et al [299] reduced graphene oxide that was created using Verbena officinalis as a reducing agent, Graphite and the modified Hummer process were used to create GO.Cross-linking was used to encapsulate rGO in sodium alginate SA to produce SA-rGO beads. Olive mill effluent has been treated with SA-rGO beads. This effluent contains significant amounts of hazardous phenolic chemicals, which prevent biological decomposition.

# 8 Applications of graphene

## 8.1 Graphene oxide (GO) and Graphene (G) in energy applications

The worldwide difficulties of energy consumption and preservation of the environment are significant concerns, with fossil fuels serving as the major source of energy [300] .Sustainable and renewable energy sources for example: Organic solar cells, capacitors, and lithium batteries, are being researched as alternatives to traditional energy sources. GO, a single layer of graphite oxide, offers remarkable features, including adjustable electrical properties, outstanding optical and mechanical characteristics, and a wide range of uses, as shown on **Table. 5** [301]. .However, GO has notable drawbacks for practical uses, including structural flaws, weak dispersion, restacking, and layered thickness [302]. The insulating properties of ordinary GO limit its use in electrical devices and storage of energy [303, 304]. Oxygenated groups can broaden the structural/chemical variety of GO by additional Chemical modifications or addition of functionality, providing an effective technique to tune its features to predicted. extents. GO-based composites and GO have shown tremendous capacity for energy storage/conversion and environmental protection. Recent advances in GO-based materials for Li batteries, Capacitors, fuel cells, and solar cells.

### 8.1.1 In electrochemical energy storage and energy conversion.

Graphene oxide (GO) material is flexible, having good characteristics and structural variety, which make it suitable for a wide range of energy storage applications [305, 306]. GO-based thin films are manufactured using conventional processes such as drop casting, spin coating, spray coating, or dip coating. Because of its amphiphilicity, GO can self-assemble on different surfaces, allowing the production of controlled microstructures and tunable characteristics [307].

GO has been researched for its possible use in electrochemical energy storage and conversion devices such as batteries, capacitors, and fuel cells, respectively. The oxygen functional groups of GO can behave as oxygen donors. In contrast, the negatively charged groups attract positively charged species for quicker ion transport and repel ions with the same charge. GO has adjustable interlayer spacing and a large specific surface area to enable electrochemical reactions while managing product size [308]. Managing product volume or size variations. GO can also function as insulating dielectric separators or electrically conducting substrates.

### 8.1.2 in batteries applications

S.J. Richard Prabakar et al. [309] The study discusses the use of graphene oxide as a protective layer to prevent corrosion of aluminum current collectors in lithium-ion batteries. A spin coating process was used to coat graphene oxide on aluminum, which acts as an excellent corrosion inhibitor in the LiPF6 electrolyte. Experiments showed that GO adheres tightly to aluminum and efficiently prevents oxidation by acting as a barrier to charge transfer to pits. The LIB comprising LiFePO4 and GO-coated Li showed improved cycle stability and performance rate compared to its bare lithium equivalent [310].

Haiwei Wu et al. [311], demonstrated that the oxide of graphene (GO)-based sulfur anode materials with high electrochemical properties have been widely documented. A unique graphene oxide (GO)/sulfur (S) hybrid is generated using the approach of electrostatic self-assembly [312, 313] Interestingly, the core-shell composition of the hybrid and the GO support assist in delaying the spreading of polysulfides through the process of the electrochemical cycle [311]. The GO/sulfur cathode improved cycling ability. Specific discharge capacities of up to 494.7 mAh g1 greater than 200 cycles at 0.1 C are attained with an improved columbic efficiency of roughly 95%, making it a suitable cathode material for lithium-sulfur batteries [314].

Peng Xian Han et al. [315] Graphene oxide nanosheets have been found to be an effective electrochemical active material in vanadium redox flow batteries for V2+/V3+ and VO2+/VO2+ redox couples [316-321]. Experimental results showed that a small amount of GO significantly improved the reversal between V3+ and V2+ and enhanced peak current for both redox reactions compared to graphite [315]. This was due to the presence of an excess of oxygen groups in GO and its high active surface area, which allowed it to act as an electrocatalyst throughout the redox process of vanadium species. GO sheets were attached to multi-walled carbon nanotube surfaces to create an electrocatalytic composite with an efficient heterogeneous conductivity network [322, 323].

Turgul Cetinkaya et al. [324], found that a Li-air battery with a graphene oxide cathode had a charge capacity of 585 mAh g1 and a discharge capacity of 612 mAh g1 after 10 cycles. The GO sheet's unique structure allows for varying porosity for oxygen transport, increasing Li2O production and cracking efficiency. A hierarchically porous carbon was synthesized from GO gel in nickel foam without external cross-linkers, providing a three-dimensional framework. The specific capacity of the Li-air cell using GO-derived porous carbon cathode reached 11,060 mAh g1 at 280 mAh g1 [325].

The manufacturing of reduced graphene (RGO)-based Co3O4 hybrid as anodes for lithium-ion (Li-ion) batteries to prepare Co3O4/RGO by microwave irradiation method [326, 327]. It has greatly affected their performance as electrodes in LIBs which graphene has a high surface area, high electrical conductivity, and good mechanical properties [328]. RGO/Co3O4 composites exhibit generally superior electrochemical characteristics as material for anodes for LIBs at high temperatures of 100°C. it is showcasing higher reversible capacities and excellent abilities after 100 cycles at 100 mAg−1, accompanied by a significant rate of performance Because of low cost and superior electrochemical efficiency[328, 329].

Yaprak Subasi et al. [330] developed a composite anode for Li-ion batteries using TiO2 NP and (RGO) composite as conductive carbon sources. A unique surface modification was performed using $H_2O_2$ to enhance the electrochemical performance of the nanocomposite anode (TiO2/RGO-P composite) [331]. The improved and stable development of 3D channels results in larger reversible capacities and better cycle performance. The TiO2/RGO-P composite anode had a reversible discharge capacity of 291 mAh g−1 at 100mAg−1, with a Coulombic efficiency >99% and an average discharge capacity loss of ~0.31% per cycle for the first 100 cycles [332]. TiO2 was chosen due to its high theoretical capacity, low cost, chemical stability, workable nanostructures, and low toxicity [332, 333].

### 8.1.3 In capacitors application

Chao Wu et al. [32] A hybrid CNT coated with GO has been developed for excellent performance in dielectric capacitors. This technology, fabricated through p-p interactions between negatively charged GO and positively charged CNTs, offers a high dielectric constant, low loss, higher breakdown strength, and better energy storage density. The GO shells promote CNT dispersion, act as protective barriers, and enhance charge dispersion due to its thin structure [334, 335]. The dielectric constant of the capacitor for hydrated GO sheets was responsive to both the water content and the applied voltage.

Guixia Zhao et al. [336] demonstrated that graphene oxide (GO) can be used as an active material in supercapacitors as electrodes. GO can serve as a template for stacked MnO2 nanosheets, increasing surface area and potentially improving capacitance performance [336-340]. Oxygen-containing GO clusters act as anchoring sites, allowing in situ production of MnO2 nanoparticles that adhere to GO sheets, enhancing surface area and preventing aggregation [337]. A GO/Mn3O4 hybrid thin film demonstrated a specific capacitance of 344 F g1 at 5 mV s1. The capacitive electrochemical behavior of graphene and graphene oxide was also documented [340, 341].

Chikako ogata et al. [342] developed a thin, flexible solid-state GO supercapacitor (SC) using GO's remarkable proton conductivity. The GO film was UV-cured to generate rGO, maintaining the GO interfacial layer in the center. This system utilized GO's exceptional ion conductivity and dielectric properties as a separator and solid-state electrolyte. A solid-state supercapacitor with cross-linked PVA/graphene hydrogel electrodes and ion-doped GO gel as a gelatinous polymer electrolyte and separator was also described. The supercapacitor with 20 wt% GO-B-PVA/KOH gel electrolyte showed consistent cycling behavior and a larger discharge capacitance [343].

Lou et al. [344] Lu et al. fabricated a NiCo-MOF/rGO hybrid material using a one-step room temperature precipitation method. MOF materials are popular for energy storage and supercapacitors due to their large surface area and high atom-to-surface volume ratios. The hybrid material produced an impressive specific capacitance of 1553 F/g at a current density of 11 A/g, demonstrating a large rotational capacity after 5000 cycles and excellent energy densities in the assembled asymmetric device [345, 346].

Gomaa A. M. Ali et al. [347] developed a composite $MoS_2$/graphene from bulk $MoS_2$ and graphite rod using a simple electrochemical exfoliation technique. The composite shows low electrical resistance, improved electrode specific capacitance, and high electrochemical stability [348, 349]. Its synergistic effect enhances charge storage due to its high specific capacitance of 227 F g−1, higher than that of graphene and MoS2 at a current density of 0.1 A g−1 [350]. The MoS2/graphene hybrid also has a long cycle life of 89% after 2000 cycles, making it a potential electrode for high-performance supercapacitors [350-352].

### 8.1.4 In fuel cell applications

Ju Hae Jung et al. [353] The study uses Pt/Graphene oxide for a polymer electrolyte membrane fuel cell. A small amount of GO enhances the durability of commercial Pt/C catalysts without compromising the initial electrochemically active surface area. GO provides anchoring sites for dilute metal ions, preventing Pt agglomeration. It can also be used as an electron donor to reduce noble metal ions without additional reductants and surfactants. This results in the growth of metal nanoparticles on GO surfaces. Incorporating GO into the polymer matrix enhances conductivity and reduces membrane contamination [354-358].

Hikaru Tateishi et al. [359] have explored graphene oxide as a potential membrane/separator in fuel cells due to its high proton conductivity [360-362]. GO has shown greater water absorption, improved water retention, and higher modules of elasticity and tensile strength than NPC membrane. The free-standing GO membrane in PEMFC showed the highest power density of ~34 mW cm2 within 30-80°C, similar to Nafion-based fuel cells [360]. In addition, Functionalized GO sheets, like ozonized GO and sulfonic acid-treated GO, show increased proton conduction due to higher oxygenated functional groups, enhancing fuel cell function [363, 364].

The preparation of Graphene-based low-temperature fuel cell technology has benefits, including a reduction in emissions associated with electricity generation, a reduction in local air pollution, and an increase in energy security [365]. Graphene has conductive properties and a high surface area that enables it to be used as an electrocatalysts for (ORR) and fuel oxidizer. It is possible to add catalysts to graphene, such as high tensile strength, a polymer that has high ionic conductivity, and low fuel permeation [366].

Sümeyye Dursun et al. [367] found that Graphene (G) (CVD), multi-walled carbon nanotube (MWCNT) [368] and Carbon black (Vulcan XC72) [369] and supported for Cobalt (II) Phthalocyanine catalysts (Co/Pc ) were prepared by impregnation technique to be used as polymer electrolyte for the membrane of fuel cells (PEMFC) and electrocatalysts for oxygen reduction reaction in acid electrolytes. Co/Pc with CVD graphene created an impermeable layer structure which resulted in lower performance and mass transport limitation. The enhancements of the Performance were achieved by operating at high temperatures, higher catalyst loading, heat treatment of the samples, and higher pressures under 100% humidity [370]. hybridization of CVD graphene with MWCNT and Vulcan led to improved fuel cell performance at 25 psi back pressure compared to individual support materials. Scientists studied three different temperatures (800 °C, 650 °C, and 500 °C) were studied to examine the effect of heat treatment on Co/Pc. With examination, the best fuel cell performance was obtained at 800 °C. Polymer electrolyte membrane fuel cells (PEMFCs) were chosen because they give zero-emission and high efficiency [371, 372].

**8.1.5 In Renewable Energy Applications**

Graphene oxide (GO) has garnered significant interest in the field of solar cells due to its novel properties. GO can serve as a transparent conductive electrode material. Its transparency and high electrical conductivity make it an excellent replacement for conventional materials like indium tin oxide (ITO) in solar cells. By coating GO on glass or flexible substrates, researchers have developed transparent electrodes that allow light to pass through while efficiently conducting electricity [373] .GO can be incorporated as a counter-electrode material in DSSCs. Its large surface area and excellent electrical conductivity enhance charge transfer kinetics, improving overall cell efficiency [373] .GO has been explored as a hole extraction layer (HEL) in OSCs. Its properties, such as high electrical conductivity and tunable work function, contribute to efficient charge extraction from the active layer, leading to better device performance [374]. In this review, the OSCs and DSSCs will be discussed below.

Jae Hoon Kim et al. [375] used graphene oxide as a hole extraction layer for stable organic solar cells due to its band gap tunability,[376-379] , high transparency, and dielectric nature [375, 380-382]. The researchers found that GO had good surface qualities and exceptional optical features with high transmittance [378]. The researchers found that GO had good surface qualities and exceptional optical features with high transmittance. The initial performance of devices containing PEDOT: PSS and GO showed almost comparable values. The enhanced lifetime was due to varying degrees of space-limited space charge sharing depending on the HELs. The impedance spectroscopy method confirmed this, with a doping ratio of 5 wt% resulting in a 24% improvement in PCE. This research will provide insights into developing high-performance HEL materials for various optoelectronic applications.

Reduced graphene oxide (rGO) was prepared from graphene using a chemical reduction method, forming monolayer sheets [383, 384]. It was used to improve the performance of dye-sensitized solar cells in a photovoltaic code. N719/rGO showed an efficiency of 2.02%, while N719 had a efficiency of only 1.60% [385]. The use of rGO in dye-sensitized solar cells improved voltage and current characteristics and electrical power, with different ratios affecting these characteristics [386].

V. Paranthaman et al. [387, 388] found that the counter electrode (CE) plays a crucial role in the electrocatalytic reaction in dye-sensitized solar cells (DSSC) to achieve maximum power conversion efficiency (PCE). They analyzed the electrocatalytic properties of RGO/NMP, RGO/NMP/PVP, and RGO/Nafion based on CE materials, measured maximum current density, and performed cyclic voltammetry

[389]. The CE-based RGO/NMP/PVP formulation showed superior electrostimulation capabilities, with a maximum PCE of 5.8%. The synthesized results showed better results for RGO/NMP/PVP, RGO/NMP, and RGO/Nafion-based DSSC [390, 391].

The synthesis of an RGO-incorporated GO host was used to create an RGO/GO nanocomposite photoelectrode using a simple solution impregnation technique [392]. RGO/GO nanocomposite photoelectrode exhibited improvement in Photoelectrochemical PEC water reduction activity and had the highest photocurrent of − 61.35 µA/cm2 while 42.80 µA/cm2 for only GO photoelectrode. RGO demonstrated a quicker electron acceptor and transporter role, as well as changing the electronic band structure (band gap) of GO, which is important for enhancing the optoelectronic capabilities of layered semiconductor materials and conjugated molecules [393].

B. Koo et al. [394] discovered that Reactive Green Oxide (RGO) enhances the photochemical stability of Cu(In,Ga)Se_2 (CIGS) photocathode for solar water splitting [395, 396]. The introduction of rGO to the photocatalyst layer (PEC) preserves Pt, enhancing stability. RGO also acts as an adhesive interlayer, diffusion barrier, efficient charge transfer layer, and protective layer. When rGO was deposited as an interlayer between CIGS/CdS and Pt, it showed the lowest R_ct of the cell, indicating optimal conditions for effective PEC performance. This is due to enhanced charge transfer via efficient lateral distribution of photogenerated electrons by rGO conductive to Pt [394].

**Table. 5** Example for enhancement Graphene in energy applications.

| Material | Method | Enhancement | Ref. |
|---|---|---|---|
| **Graphene oxide (GO)** | Electrostatic self-assembly | Electrochemical performance for rechargeable lithium-sulfur batteries. | [311] |
| **Sulfur-Graphene Hybrid** | Two-Step Hydrothermal Method | Electrochemical performance for lithium-sulfur batteries | [313] |
| **Reduced graphene oxide with tunable C/O ratio** | | GO electrochemical activity towards $VO^{2+}/VO^{2+}$ and $V^{3+}/V^{2+}$ redox couples are tunable which enhances. | [320] |
| **the MnO2-GO hybrid with ($MnO_2$) nanosheet arrays on graphene oxide (GO) flakes.** | simple hydrothermal method | The performance of vanadium redox flow battery. Advanced electrode material for supercapacitors | [340] |
| **N719/rGO** | ----------------- | The performance and efficiency of the dye-sensitized solar cell and improvement of electric power to the solar cell. | [386, 391] |
| **CIGS/CdS/rGO** | ----------------- | To enhance the Photoelectrochemical Stability, charge transfer PEC for efficient PEC performance. | [394] |
| **RGO/GO Nano-hybrid photoelectrode** | Simple solution impregnation method | Improvement in Photoelectrochemical PEC water reduction activity, exhibited a transporter role and faster electron acceptor, influencing (band gab) of Graphene oxide | [393] |
| **$Co_3O_4$/RGO** | microwave irradiation method | Electrodes in lithium-ion batteries | [329] |
| **NiCo-MOF/rGO hybrid** | room temperature precipitation process | Supercapacitor application and energy storage | [391] |
| **$MoS_2$/Graphene composite** | electrochemical exfoliation method | To improve the specific capacitance of the electrode | [350] |
| **Graphene oxide** | The Hummers method. | Hole extraction layer for stable organic solar cells | [375] |
| **Pt/graphene oxide and Pt/C hybrid catalyst** | Hummer's method | Polymer electrolyte membrane fuel cell. | [353] |

## 8.2 Graphene in electrical Applications

Graphene, a one-atom thick material, is revolutionizing electrical applications due to its high electrical conductivity, thermal stability, and mechanical strength [397]. Its ability to attenuate electromagnetic waves protects electronic equipment, while its conductive nature allows for more durable and efficient RFID tags. Its high surface area and sensitivity to environmental changes enable the development of accurate and fast sensors. Graphene's flexibility and conductivity are also used in screen-printed applications, paving the way for innovative designs in flexible electronics and wearable devices, as shown on **Table. 6** [1, 398].

### 8.2.1  In Electromagnetic interference shielding (EMI SE) applications.

Seul Ki Hong et al [399, 400] fabricate single layer graphene by chemical vapor deposition (CVD) process for inductively coupled plasma. They used graphene due to its saturation velocity, flexibility, mechanical strength, and its high conductivity. Since the average values of the absorbance, reflectance, and SE are −4.38 dB, −13.66 dB and 2.27 dB respectively. According to these results shielding for incident waves can be about 40% by single layer graphene that is synthesized by CVD. The command mechanism of the EMI SE of the single layer graphene is absorption before reflection.

Enhui Zhu et al. [401] successfully fabricated graphene aerogel (GAs) by hydro plastic foaming (HPF). Extreme-stable HGA is fabricated with outstanding EMI SE performance which can reach 64.1 dB at density of 3.7 mg cm$^{-3}$ and the thickness of 1 mm. The specific shielding come to 173,243 dB cm$^2$ g$^{-1}$ [402], which is the uppermost between earlier carbon-based material that lightweight, as the abundant interfaces formed in the elevated porosity structure of graphene aerogel leads to many reflections of EM waves to absorb more energy.

Marta Gonzalez et al. [403]  were able to synthesize graphene aerogels with reduced and controlled porosity at different elevated temperatures by an adjusted hydrothermal treatment. The specific SE efficiency of the samples synthesized in this mission reaches up to 6743 dBcm$^3$ g$^{-1}$ (for GR:1000 °C  sample) and 2091 dBcm$^3$ g$^{-1}$ (for GR sample). Electromagnetic characterization has detected that the thermal treatment temperature when changed can modify the electromagnetic mechanisms. As the heat treatment temperature increases, the reflection strength increases due to the increase in the reduction efficiency, which increases the electrical conductivity. With the increase in the capability of the material to reflect incident radiation, this leads to a decrease in the sum of radiation that enters the material.

Chaobo Liang et al.[404] were fabricated RGFs/epoxy resin (RGFs/EP) for EMI SE based on epoxy resin (EP) and reduced graphene oxide films (RGFs) that synthesized by the bar-coating method subsequently chemical and thermal reduction by inserting several layers of RGFs into the epoxy resin matrix by a pre-arrangement method. Insert of extremely compatible RGFs with layered structure was extra suitable for benefiting the EMI shielding performance (SE) and in-plane electrical conductivity of RGFs/EP EMI SE combinations as electromagnetic waves are hold and mitigated by reflection, several scattering and absorption within RGFs with great area reflecting surface, giving rise to high values superb EMI shielding [405]. An outstanding EMI shielding value of 82 dB was accomplished for 8-RGFs/EP EMI SE combinations, roughly 41 times greater than that of pure epoxy resin (∼2 dB) at X-band. The SE mechanism is majorly imputed to absorption. Meanwhile, the acquired in-plane electrical conductivity excesses pointedly from $10^{-12}$ S/m of pure epoxy resin to 591.7 S/m of 8-RGFs/EP EMI shielding combinations. The corresponding THRI (Heat Resistance Index) value of 8- RGFs/EP EMI shielding hybrids excesses from 183.0°C (pure epoxy resin matrix) to 189.5°C, exhibiting enhanced thermal stability.

Pravin Sawai et al.[406] were fabricated rGO filled PEI film/nanocomposites that based on Polyetherimide (PEI) film and Reduced Graphene Oxide (rGO) that manufactured by Hummer's method [407]. Results show that as the frequency increases, the shielding effectiveness (SE) increases as the shielding of the composites depends on the frequency and as the filler loading (rGO) in the PEI matrix increases, the electrical conductivity and SE increase and the thermo-mechanical properties improve. At 8 to 12 GHz the nanocomposites-built matrices with 2.5 wt % freight of rGO showed SE levels between 22-26 dB compared to the pure polymer (PEI) which exhibited value varying between 1-3 dB. With the increase of the wt % of rGO concentration in the polymer, the PEI-rGO nanocomposites exposed better that the specific storage modulus and thermal stability of the PEI-rGO film increased pointedly.

Bin Shen et al [408, 409] prepared a graphite oxide (GO) film by preparing a suspension from graphite oxide in H2O[410]. The film had a thickness of about 1.1 nm, indicating a monolayer characteristic. The film was made by forcing the GO into Teflon dishes and allowing H2O to evaporate under suitable conditions. The EMI SE performance of the GF-2000 sample was tested by bonding it onto a polyolefin elastomer (POE) substrate with double-sided tape. The EMI SE of the GF-2000 sample was found to be about 9.1 dB, with a higher value of about 19.1 dB at a thickness of about 8.4 μm. This Sheild Performance value is approximately the target value for practical application.

Dengguo Lai et al.[411] are successfully prepared porous graphene films (PGFs) to achieve EMI Shielding. The structure has outstanding foldability and stability due to the homogeneous fine voids introduced by PGFS and thus the elastic properties didn't reduce after folding for 3000 cycles. Due to the layer-by-layer structure of graphene, it possesses ultra-thin thickness and super-conductivity [412], thus when covering X-band frequencies LPGF offers outstanding EMI SE capacity of 43.8 dB with absorption-dominant shielding advantage. LPGF has highly graphene SE efficiency that the highest among fabricated shielding materials that are built on graphene produced to date. The LPGF film has a remarkably high specific shielding/thickness of up to 29178 dB cm$^2$ g$^{-1}$ due to the low thickness of 200 mm and low density (∼ 75 mg cm$^{-3}$).

Chi Fan et al.[413] created a flexible multilayer graphene film (MGF) with a small thickness and high conductivity using high-temperature heat treatment and compression coiling. The MGF film had a high signal-to-noise ratio (SE) of 70 dB from 2.6 to 40 GHz and 80 dB with a thickness of 54 μm. The first MGF-based multibeam millimeter wave antenna array was achieved with a 120° beam sweep range and average gain of 15.07 dBm.

## 8.2.2 In Radio-frequency identification (RFID) applications

Bohan Zhang et al. [414] designed flexible, long-range ultrahigh frequency (UHF) RFID tag antenna made of graphene-built film synthesized through thermal treatment of organic polyimide precursors[414, 415]. Tags are synthesized and experienced utilizing the fabricated antenna, which are revealed to have achieved radiation efficiency exceeded 90% in the whole ultrahigh frequency band from 860 to 960 MHz and gain above −1.5 dBi. The suggested tag reading range is greater than 12.3 m over the whole ultrahigh frequency band with an ultimate value of 14 m at 920 MHz. The read range only reduces by 4.5 m compared to the frist state at 915 MHz Over time, after 2000 cycles of stretching and bending the flexibility of the tags proven while the classical metallic tag would break.

Bohan Zhang et al. [416] fabricated and measured UHF RFID tags with antennas made of a highly flexible graphene assembly film are characterized by low sheet resistance of 0.024 Ω sq$^{-1}$ and elevated conductivity of 1.6* 10$^6$ S m$^{-1}$.[417, 418]Results of measurements show that the measured radiation efficiency surpassing

90% above the whole ultrahigh frequency band, achieved gain keeps above 2.5 dBi across the whole ultrahigh frequency band (from 860 MHz to 960 MHz). At 910 MHz the peak achieved gain can be up to 1.6 dBi and the extreme distance for continued reading state with an outdoors handheld reader is 11 m. GAF has mechanical flexibility so it can hold thousands of curvature cycles with a slight variate in resistance (It was concluded that the conductivity of GAF keeps stable before, after and during bending, because the conductivity of the material is specified by thickness and the electrical its resistance).After 500 cycles the reading range will be decreased by just 1.5m from (12m to 10.5m) and the reading range is still 8.5 meters even after 2000 cycles.

Mitra Akbari et al.[419, 420] managed to fabricate graphene-based dipole antennas using doctor-blading method [421]. The graphene antenna's measured sheet resistance is 1.9 Ω/sq. At 889 MHz, a planar dipole antenna made of graphene with a length of 143 mm realized a gain of –2.18 dBi and a recorded overall efficiency of 40%. Furthermore, at 950 MHz, a passive ultrahigh-frequency radio-frequency tag with a cardboard graphene dipole antenna was able to obtain a read range of more than five meters.

Bohan Zhang et al. [422] used high-conductivity graphene assembly film (HCGAF) to create an anti-metal flexible RFID tag antenna. Three procedures were used to create HCGAF: static calendering, high-temperature heat treatment, and graphene oxide suspension self-assembly. Static calendering was used to attach the HCGAF to a Polyethylene Terephthalate substrate following annealing. The HCGAF showed a higher conductivity reach to ($1.82*10^6$ S m$^{-1}$) by using the Four-Point Probes Resistivity Measurement System. Using SEM shows that skin depth of e HCGAF at 900 MHz was 12 μm; the HCGAF's thickness had sufficient space for electrical current flux. The HCGAF also shows an excellent mechanical flexibility that it bent from flat to 90º thousands of times without fracture.

### 8.2.3 In sensors applications

Farheen Khurshid et al.[423] were verified a sensing device based on graphene oxide (GO) that fabricated by Hummer's method [424]to detect ammonia gas. The study was performed under different concentrations and different humidity provisions of ammonia, ranging from 100 to 400 ppm. Results show that as the $NH_3$ concentration increased from 100 to 400 ppm, the ammonia gas sensitivity of GO enhanced from 45 to 58% and as the humidity level enhanced from 11% to 75% the sensitivity of GO enhanced from 36 to 68%. The interaction energy for adsorption of ammonia by diverse functional groups like (hydroxyl and carboxyl) on GO is 37 kJ/mol and 46 kJ/mol, respectively.

Ruma Ghosh et al. [425, 426] were fabricated reduced graphene oxide sensor (RGO) to detect ammonia gas and conducted several tests for diverse concentrations of $NH_3$ gas. From testing diverse concentrations of $NH_3$ gas, it was found that with increasing ammonia concentration the answer of the RGO sensor enhanced (as the response ranged from 5.5% at 200 ppm to 23% at 2800 ppm $NH_3$). From the different reduction times, compared to $NH_3$ gas sensors built on carbon nanomaterials, it was found that the response of reduced RGO at 90 minutes was much better.[427] The RGO sensor was found to be more selective towards ammonia compared to different volatile organic compounds (VOCs).

Chang Wang et al. [428] were fabricated Reduced GO-Graphene (RGO-Gr) combined gas sensor that based on pure graphene that synthesized by a chemical vapor deposition method[429] ornamented on RGO, for very-low concentration ammonia revelation to take advantages of the great number of adsorption locations found in RGO and the elevated conductivity of graphene [430]. They also studied the influence of RGO reduction time on the sensing properties in order to improve the states of these gas sensors. The superior gas sensing response demonstrated in this kind of sensor is the RGO-Gr gas sensor with a reduction time of 10 minutes up to 14.67%. The detection limit of this gas sensor is 36 ppb, and the highest sensitivity towards 0.5 ppm is 2.88%. Compared with the RGO gas sensor, it significantly improves the anti-noise capability with the signal-to-noise ratio (SNR) increased from 22 to 1008. RGO on graphene combined structure has robust selectivity toward ammonia and perhaps because of the robust reduction and donor behavior of ammonia between the inserted vapors.

Somayeh Tohidi et al.[431] fabricated PANI/3D-RGO hybrids as NH3 gas sensors using electrochemical deposition methods. The integrated sensor is highly selective for NH3 upon exposure to various gases. The combined sample's surface area is significantly higher than the pure PANI sample, at 40.544 m2 g-1. The high surface area of 3D/RGO is beneficial for gas absorption and desorption. The three-dimensional porous structure of 3D-RGO is suitable for gas diffusion and adsorption due to its disorder and high energy binding sites and edges [432, 433] The PANI/3D-RGO hybrid sensor is the best solution for 50 ppm ammonia gas at room temperature and lower concentrations. The response time of the combined PANI/3D-RGO sensors is reduced compared to pure PANI sensors due to the high carrier mobility available from the RGO layers.

Rafaela S. Andre et al. [434] have developed a NFI-rGO composite sensor, which combines reduced graphene oxide (rGO) and ceramic nanofiber $In_2O_3$ (NFI) [435] he sensor has a rapid response with 10 times higher sensitivity than rGO and NFI individually, a response time of 17 seconds, and a low detection limit of 44 ppb. It also shows great selectivity for anti-ammonia organic solvents and other nitrogenous compounds. When exposed to $NH_3$, the gas molecules react with rGO absorption centers and NFI absorbed oxygen species, reducing total resistance. The hybrid improvement is due to the combination of NFI and rGO, increasing active sites for $NH_3$ gas adsorption and facilitating gas access to additional adsorption centers [436, 437].

### 8.2.4 In screen printed applications.

Miguel Franco et al.[438] developed a capacitive multitouch sensing surface by conductive graphene nanoparticles-built ink with carboxymethyl cellulose as a binder. They used the eco-friendly 90 GNP/CMC ink which developed a 40 columns × 28 rows 8" with a circuit resistance of 400 kΩ for the top and 151 kΩ for the bottom.

Dong Seok Kim et al.[439] created a highly conductive and concentrated ink built on disorder-free graphene using a scalable fluid dynamics method. For graphene ink, mixing and a high shear exfoliation process yielded graphene at an elevated concentration of 47.5 mg mL$^{-1}$. Even when bent physically, the screen-printed graphene conductor maintains its remarkable electrical conductivity of $1.49 \times 10^4$ S m$^{-1}$.

Hui Ding et al [440] managed to print an aqueous highly conductive graphene ink on a paper substrate utilizing screen printing technique. The fully controlled rheological characteristics of the graphene ink allow for high resolution printing of lines up to 45 μm in width. The printed lines of graphene exhibit an elevated conductivity of reach to $4.23 \times 10^4$ S m$^{-1}$ at 3 μm-thick after being roller crushed and dried at 70 °C. Additionally, it has outstanding curvature stability throughout the roller stress process, making it suitable for industrial production.

Pei He et al.[441] succeeded in fabricating highly conductive graphene patterns on paper and flexible plastic substrates by utilizing a screen-printing process. Following post annealing and compression rolling, the screen-printed graphene patterns on PI substrates demonstrated a high electrical conductivity of around $8.81 \times 10^4$ S m$^{-1}$. Suitable for flexible electronic applications, the printed graphene patterns demonstrated exceptional mechanical steadiness on both paper and plastic substrates. Solid-state supercapacitors were successfully built on flexible substrates and demonstrated an area capacitance of around 40 mF cm$^{-2}$, together with good rate ability and flexibility, using highly conductive graphene patterns as the present collector.

Xiao You et al [442] succeeded in fabricating one-dimensional graphene-based fibers by direct ink extrusion. The graphene-based fibers are capable of twisting, stretching, curvature, and compression in addition to other deformations. Multiple strain cycles can yield a variety of response amplitudes with a reversible electrical resistance change, demonstrating the great sensitivity and wide domain of strain sensing. The interwoven graphene network is responsible for the ultra-sensitive electromechanical feature, which has a mensuration factor of 65 under 6% strain and can withstand 600 cycles of continual stretching-releasing operation with less than 6.2% feebleness in the answer signal.

**Table. 6** Example for enhancement Graphene in electrical applications.

| Material | Method | Enhancement | Ref. |
|---|---|---|---|
| **Monolayer Graphene** | *Inductively coupled plasma (ICP) chemical vapor deposition (CVD)* | The Electromagnetic interference Shielding Effectiveness of monolayer graphene | [399] |
| **Aerogel Graphene** | Hydroplastic foaming (HPF) | The Electromagnetic interference Shielding Performance | [443] |
| **RGO Graphene** | *The bar-coating method followed by chemical and thermal reduction* | The Electromagnetic Interference Shielding Performance | |
| **Ultrathin Flexible Graphene Film** | -------------------------- | The Efficiency of Electromagnetic Interference Shielding | [408] |
| **Graphene based RFID tag antenna** | *Thermal treatment of organic polyimide precursors* | The Radiation Efficiency of RFID tag antennas | [444] |
| **High-conductivity graphene assembly film** | ----------------------------- | The conductivity and mechanical flexibility of RFID tag antennas | [445] |
| **Graphene oxide (GO) based sensor** | *Hummer's method* | Increasing of sensitivity and humidity level of the ammonia gas of GO | [446] |
| **Reduced graphene oxide (RGO) based sensor** | ------------------------- | Increased response of RGO to ammonia concentration | [448 ,447] |
| **graphene-based ink** | *Scalable fluid dynamics method* | Increasing conductive and physical properties of graphene ink | [449] |
| **graphene-based fibers** | *Direct ink extrusion* | The properties of graphene-based fiber ink | [450] |

## 9. Conclusion

The exploration of 2D graphene material within this study has illuminated its remarkable potential across a spectrum of applications. Graphene's unique properties, such as its exceptional electrical conductivity, high mechanical strength, and superior thermal properties, have been harnessed to revolutionize energy solutions. Its deployment in batteries, energy storage systems, capacitors, fuel cells, and renewable energy technologies underscores a transformative shift towards more efficient, durable, and compact energy devices. Furthermore, graphene's role in electronics—from RFID tags to sensors and EMI shielding—demonstrates its versatility and capacity to enhance performance and miniaturization. This study not only shows cases graphene's current applications but also sets the stage for future innovations that could redefine the technological landscape. As we stand on the cusp of a materials revolution, graphene emerges as a cornerstone, promising to uphold and advance the integrity of our energy and electronic frontiers.